\documentclass[letterpaper]{JHEP3}
\pdfoutput=1

\usepackage{graphicx}
\usepackage{amsmath,amsthm, amssymb}

\newcommand{\eq}{\begin{equation}}
\newcommand{\eqe}{\end{equation}}

\newcommand{\eqa}{\begin{eqnarray}}
\newcommand{\eqae}{\end{eqnarray}}

\newcommand{\e}{\epsilon}

\title{Amplitudes of 3D and 6D Maximal Superconformal Theories in Supertwistor Space.}
\author{Yu-tin Huang\footnote{Email: yhuang@physics.ucla.edu}$^{~1}$, Arthur E. Lipstein \footnote{Email: arthur@theory.caltech.edu}$^{~2}$
\\ \\
\\
\it $^1$ Department of Physics and Astronomy,\\ UCLA,\\
Los Angeles, CA 90095-1547, USA
\\ \\
\\
\it $^2$ California Institute of Technology,\\
Pasadena, CA 91125, USA\;}
\abstract{We use supertwistor space to construct scattering amplitudes of maximal
superconformal theories in three and six dimensions. In both cases, the constraints of superconformal invariance and rationality
imply that the three-point
amplitude vanishes on-shell, which constrains the four-point amplitude to have vanishing residues in all channels.
In three dimensions, we find a unique solution for the four-point amplitude and demonstrate that it agrees with the component result
in the BLG theory. This suggests that BLG is the unique three-dimensional theory with classical $OSp(8|4)$ symmetry that admits a Lagrangian
description. We also show
that one can derive the four-point amplitude of the ABJM theory from our $\mathcal{N}=8$ result by
reducing the supersymmetry, which implies that the tree-level Yangian
symmetry recently found in ABJM is also present in BLG. In six
dimensions, we find that the consistency conditions imply that all
tree-level amplitudes vanish. This leads us to conjecture that an interacting six-dimensional theory with classical $OSp(8|4)$ symmetry does not have a  Lagrangian description, local or nonlocal, unless
the $(2,0)$ tensor multiplets are supplemented by additional degrees of
freedom.
\;
}
\preprint{CALT 68-2787, UCLA-TEP-10-104}

\begin{document}

\numberwithin{equation}{section}
\section{Introduction}
In recent years, the study of scattering amplitudes of four dimensional (super)Yang-Mills and (super)gravity has revealed many surprises. These include new symmetries, structures~\cite{Drummond:2008vq, Witten:2003nn,Cachazo:2004kj,Britto:2004ap} and dualities~\cite{Bern:2008qj,Mason:2009sa, ArkaniHamed:2009dn}. These results are at times obscure from the point of view of the action, and only become manifest when one considers on-shell objects such as S-matrix elements. Many of these advances are largely due to the spinor-helicity formalism which uses (super)twistors to covariantly parameterize on-shell momenta, polarization vectors, and (for supersymmetric theories) on-shell multiplets. The progress that has been made in four dimensions inspires one to look for hidden structures in the amplitudes of massless theories in $D\neq4$ dimensions.

In four dimensions, the supertwistors are in the spinor representation of the superconformal group $SU(2,2|N)$. There are two other well known cases where a supertwistor representation exists: in three dimensions where the superconformal group is $OSp(N|4)$, and in six dimensions where it is $OSp^*(8|2N) $\cite{Kugo:1982bn,Penrose:1986ca}. Recently, these (super)twistors have been used to study amplitudes of various theories in three and six dimensions \cite{Agarwal:2008pu, Cheung:2009dc, Dennen:2009vk}. In this paper, we would like to consider amplitudes of maximal superconfomal theories.

Maximal theories in three and six dimensions should describe the low energy dynamics of the world volume theory of M2- and M5-branes, which are the fundamental objects of M-theory \cite{Maldacena:1997re}. In particular, the world-volume theory for multiple M2-branes is three-dimensional with $OSp(8|4)$ symmetry and the world-volume theory for multiple M5-branes is six dimensional with $OSp^*(8|4)$ symmetry. The first three-dimensional theory with classical $OSp(8|4)$ symmetry was constructed by Bagger, Lambert, and independently Gustavsson (BLG) \cite{BLG}. This theory was based on a totally anti-symmetric 4-index structure constant which obeys a generalization of the Jacobi identity. Furthermore, it describes two M2-branes propagating in a non-trivial spacetime background. Unfortunately, it does not seem possible to generalize this theory to describe an arbitrary number of M2-branes without sacrificing classical $OSp(8|4)$ symmetry. Indeed, the theory describing an arbitrary number of M2-branes, which was discovered by Aharony, Bergman, Jafferis, and Maldacena (ABJM), only has classical $OSp(6|4)$ symmetry \cite{ABJM, Bandres:2008ry}.

Although there's been a lot of progress in constructing the world-volume theory of multiple M2-branes, the world-volume theory for multiple M5-branes
remains a mystery. A lot of progress can be made, however, if the two dimensions of the M5-branes wrap a Reimann surface with punctures giving rise to
a four-dimensional theory \cite{Gaiotto:2009we}. Since the six dimensional theory is inherently strongly coupled, a Lagrangian description may not exist
\cite{Seiberg:1996qx}, but if such a description is possible, the degrees of freedom should include $(2,0)$ tensor multiplets, each of which contains 5 scalars, 2 Weyl fermions, and a 2-form gauge field with self-dual field strength \cite{Becker:1996my}. Constructing an interacting Lagrangian
using tensor multiplets is challenging because the constraint of conformal invariance severely restricts the types of interactions one can have.
For example, a four point contact term of scalars would already have mass dimension 8 requiring a coupling constant with mass dimension -2. Furthermore,
it is difficult to construct theories of self-interacting self-dual tensor fields even without the constraint of conformal invariance. For the
world-volume theory of a single M5-brane (where all the interactions break conformal invariance), this can be achieved either by sacrificing manifest
Lorentz invariance \cite{Perry:1996mk} or by introducing an auxiliary scalar which enters the action in a non-polynomial way \cite{Pasti:1996vs}.
It is natural to try generalizing the world-volume theory for a single M5-brane to describe multiple M5-branes by introducing non-abelian group
structure, but it is unclear how to define a non-abelian anti-symmetric tensor field.

One of the biggest advantages to working with S-matrix elements in supertwistor space is that one can gain insight into a superconfomal theory even if
one does not have an action at hand. Indeed, we will try to answer the following question: assuming the existence of a not yet found action, can one
obtain information about this action by constructing superconformal invariants and imposing consistency conditions such that the invariants can be
interpreted as amplitudes? Note that this question is valid even if the coupling constant of the action is large and a perturbative expansion
is meaningless. The reason is that if one had an action, one could still naively compute tree level amplitudes obtaining rational functions of
kinematic invariants that obey the symmetries of the action. Hence, by searching for superconformal invariants that are rational functions,
one should obtain information about the structure of the unknown action, even if it is nonlocal.

The drawback of this approach, however, is that superconformal invariance may not be enough to uniquely fix all the amplitudes. Indeed for $\mathcal{N}=4$ sYM, it is only after imposing (dual)superconformal
symmetry and the correct collinear factorization that one completely fixes the tree-level amplitudes \cite{Bargheer:2009qu}. This ambiguity is not an
issue for lower-point amplitudes, however. For example, conformal invariance is sufficient to fix the three point amplitude of Yang-Mills, which is
totally anti-symmetric and thus implies that theory has a totally anti-symmetric three-index structure constant.

For the amplitudes that we consider, superconformal invariance provides very stringent constraints. For example, in three dimensions we find that all odd-point amplitudes vanish. The fact that the three-point amplitude vanishes in turn imposes an
additional constraint on the four-point amplitude, namely that the residue in all channels must vanish. This uniquely fixes the
four-point amplitude, and implies that the theory has a totally anti-symmetric four-index structure constant. Furthermore, we find that the four-point amplitude matches that of the BLG theory. These results suggest that the BLG model is the only theory with a Lagrangian that has classical $OSp(8|4)$ symmetry and is unitary.\footnote{It has been argued that the superconformal symmetry of ABJM becomes enhanced to $OSp(8|4)$ at level $k=1,2$ with the inclusion of monopole operators~\cite{Gustavsson:2009pm}. This introduces new degrees of freedom that are not included in our on-shell multiplet. The uniqueness of BLG can also be derived by coupling the vector and scalar multiplets in $\mathcal{N}=8$ superspace and requiring conformal invariance and polynomial interactions~\cite{Samtleben:2009ts}.} In six dimensions, we find that the three-point amplitude also vanishes and that we cannot construct a rational superconformal four-point amplitude which has vanishing residues in all channels. We extend this analysis to show that all tree-level amplitudes must vanish. This leads us to conjecture that a
superconformal interacting Lagrangian cannot be constructed using only $(2,0)$ tensor multiplets, although it may be possible to construct a
Lagrangian if one introduces additional degrees of freedom.

In the process of completing this paper, a paper was posted which uses three-dimensional twistor techniques to construct scattering amplitudes in the ABJM theory~\cite{Bargheer:2010hn}. In this work, the authors showed that the four and six-point tree amplitudes are invariant under a Yangian symmetry. Since the Yangian generators are built out of $OSp(6|4)$ superconformal generators, and the BLG theory has $OSp(8|4)$ superconformal symmetry, one might expect that the tree-level amplitudes of the BLG theory are invariant under a Yangian symmetry based on the $OSp(6|4)$ subgroup of $OSp(8|4)$. In fact, we will demonstrate that the four-point amplitude of ABJM can be derived from our $\mathcal{N}=8$ superamplitude, implying that the four-point amplitude of BLG also has Yangian invariance.

This paper is constructed as follows. In the next section we give a brief description of supertwistors in three dimensions and use
them to write the superconformal generators in twistor space. In section 3, we introduce harmonic variables to pick out the independent fermionic supertwistor components, which allows us to construct an on-shell superspace. We find that all of the on-shell degrees of freedom for the
three-dimensional theory can be encoded in a scalar superfield defined on this space. In section 4, we begin our construction of S-matrix elements
in three dimensions by identifying the Lorentz-invariant building blocks. We then demonstrate that a conformally invariant three-point amplitude
cannot be constructed. This leads to collinear constraints that uniquely fix the four-point amplitude. We then show that our four-point superamplitude matches
with the component result of BLG. We close this section by extracting the four-point amplitude of ABJM from our BLG result. In sections 5, 6, and 7 we
repeat these steps for six dimensions, and show that all tree-level amplitudes vanish. Section 8 presents our discussion and conclusions.
In Appendix \ref{APPA}, we review the BLG and ABJM theories. In Appendix \ref{APPB}, we define color-ordering for the bi-fundamental formulation of
the BLG theory and use it to compute the tree-level four-point scalar amplitude. In appendix \ref{APPC}, we describe some properties of $SU(2)$ spinors
which are useful for analyzing three-point amplitudes in six dimensions.

\section{3D Supertwistors and Superconformal Generators}
The supertwistors for a maximal superconformal theory in three dimensions with $SO(8)$ R-symmetry are in the spinor representation of the supergroup $OSp(8|4)$:
$$\zeta^{\mathcal{M}}=\left(\begin{array}{c}\xi^{\mu } \\ \eta^{I} \end{array}\right),\;\;\mu=1,\cdot\cdot,4, \;I=1,\cdot\cdot,8.$$
The bosonic part of the twistor $\xi^{\mu}$ is a four-component spinor transforming in the $\bf{4}$ of $SO(3,2)=Sp(4)$, while the fermionic part $\eta^{I}$ is an eight-component spinor transforming in the $8_v$ representation of $Spin(8)$.
These twistors are real and self-conjugate, i.e. they satisfy the following canonical commutation relations:
\eq
[\xi^\mu,\xi^\nu]=\Omega^{\mu\nu}\;\;\{\eta^I,\eta^J\}=\delta^{IJ},
\label{conjugation}
\eqe
where $\Omega^{\mu\nu}=-\Omega^{\nu\mu}$ and $\delta^{IJ}$ are the $Sp(4)$ and $SO(8)$ metric respectively. Note that the superalgebra metric is given by $G^{\mathcal{M}\mathcal{N}}$ where $\mathcal{G}^{IJ}=\delta^{IJ}$, $\mathcal{G}^{\mu \nu}=\Omega^{\mu \nu}$, and all other components are zero. The superconformal generators are then simply
\eq
G_{\mathcal{M}\mathcal{N}}=\zeta_{[\mathcal{M}}\zeta_{\mathcal{N})},
\label{representation}
\eqe
where one antisymmetrizes with respect to $SO(8)$ indices and symmetrizes with respect to $Sp(4)$ and mixed indices, i.e. $A^{[IJ)}=\frac{1}{2}\left(A^{IJ}-A^{JI}\right)$, $A^{[\mu\nu)}=\frac{1}{2}\left(A^{\mu\nu}+A^{\nu\mu}\right)$, and $A^{[\mu I)}=\frac{1}{2}\left(A^{\mu I}+A^{I \mu}\right)$. Using the commutation relations in eq.(\ref{conjugation}), one can show that $G^{\mathcal{M}}\,_{\mathcal{N}}$ reproduces the superconformal algebra.

The $Sp(4)$ spinors can be broken into two $SL(2,R)$ spinors
$$\xi^{\mu}=\left(\begin{array}{c}\lambda^\alpha \\ \mu_\beta\end{array}\right),\quad[\lambda^\alpha,\mu_\beta]=\delta^\alpha_\beta,\quad \alpha=1,2 $$
where the $\lambda^\alpha$'s are the usual $SL(2,R)$ spinors used for the bi-spinor representation of massless momenta:
$$p^\mu(\sigma_\mu)^{\alpha\beta}\rightarrow \lambda^\alpha\lambda^\beta.$$
Note that this gives three components since the right-hand-side is symmetric in the spinor indices. The superconformal generators can now be written in terms of 3D Lorentz indices by noting that $SO(2,1)=SL(2,R)$ and taking $\mu_\beta\rightarrow-\frac{\partial}{\partial \lambda^\beta}$:
\eqa
\nonumber P^{\alpha\beta}&=&\lambda^\alpha\lambda^\beta\quad(3)\\
\nonumber Q^{\alpha}_I&=&\frac{1}{\sqrt{2}}(\lambda^{\alpha}\frac{\partial}{\partial \eta^{I}}+\lambda^{\alpha}\eta_I)\quad(16)\\
\nonumber M^\alpha\,_\beta&=&\lambda^{\alpha}\frac{\partial}{\partial\lambda^{\beta}}-\delta^\alpha\,_\beta\lambda^{\gamma}\frac{\partial}{\partial \lambda^{\gamma}}\quad(3)\\
\nonumber D&=&\frac{1}{2}(\lambda^{\alpha}\frac{\partial}{\partial \lambda^{\alpha}}+1)\quad(1)\\
\nonumber R_{\,\,\,\, J}^{I}&=&\eta^{I}\eta_{J}-\frac{1}{2}\delta^{I}_{J}\quad(28)\\
\nonumber S_{\alpha}^I&=&\frac{1}{\sqrt{2}}(\eta^{I}\frac{\partial}{\partial \lambda^{\alpha}}+\frac{\partial}{\partial \lambda^\alpha}\frac{\partial}{\partial \eta_{I}})\quad(16)\\
\nonumber K_{\alpha\beta}&=&\frac{\partial}{\partial \lambda^{\alpha}}\frac{\partial}{\partial \lambda^{\beta}}\quad(3).\\
\label{3dgen}
\eqae
The numbers in the parentheses are the number of group elements. One can check they indeed generate the superconformal algebra,
\eqa
\nonumber \{Q^\alpha_I,Q^\beta_J\}=P^{\alpha\beta}\delta_{IJ}&,&\;\;\{S_\alpha^I,S_\beta^J\}=K_{\alpha\beta}\delta^{IJ}\\
\nonumber[D,P^{\alpha\beta}]=P^{\alpha\beta}&,&\;\;[ D, K_{\alpha\beta} ]=-K_{\alpha\beta}
\eqae
$$\{S_\alpha^I,Q^\beta_J\}=\delta^\beta_\alpha(\delta^I_JD+R^I_{\,\,\,\, J})+\delta^I_JM_\alpha\,^\beta$$
$$[P^{\alpha\beta},K_{\gamma\delta}]=-\delta^{(\alpha}_{(\delta}(\delta^{\beta)}_{\gamma)} D+M^{\beta)}_{\gamma)}).$$
The constant in the dilatation operator can be fixed by the algebra in the last line. This constant basically counts the engineering dimension of a scalar, which is $\frac{1}{2}$ in three dimensions. Note that in contrast to $\mathcal{N}=4$ sYM, the fermionic generators are linear combinations of a one-derivative operator with a zero or two derivative operator. This is due to the fact that the $\eta^I$'s are self-conjugate. In the next section, we will describe how to project out the independent fermionic components using harmonic variables.

\section{3D $\mathcal{N}=8$ On-shell Superspace}
In this section we would like to setup the on-shell superspace. This will allow us to define a superfield whose superspace expansion contains the complete on-shell multiplet, which contains eight scalars $\phi^A$ and eight two-component Majorana spinors $\psi^{\dot{A}}$. We will use conventions where the scalars and fermions are in the $8_s$ and $8_c$ representations of $Spin(8)$, respectively.

Since $\eta^{I}$ is self-conjugate, it only contains 8/2=4 degrees of freedom, which correspond to the number of on-shell supersymmetries. In order to use it to parameterize the on-shell multiplet, we need to project out its independent components. Note that this problem is similar to that of constructing an off-shell formulation for extended supersymmetric theories. There the problem is to find a set of $D^I_\alpha$ to constrain some superfield $W(x,\theta,\bar{\theta})$ (in the form of the Grassmann analyticity condition $D^I_\alpha W=0$~\cite{Galperin:1980fg}) without implying field equations through
$$\{D^I_\alpha,D^J_\beta\}=-ig^{IJ}\gamma^{\mu}_{\alpha\beta}\partial_\mu,$$
where $g^{IJ}$ is the metric for the R symmetry group. One solution is to use coset variables to project out a set of anti-commuting $D^I_\alpha$'s~\cite{Galperin:1984av}. In particular, one introduces bosonic variables $u^i_I$ which parameterize the coset $G/H$, where $G$ is the R-symmetry group acting on the index $I$, and $H$ is some subgroup (usually the Cartan subgroup) acting on the index $i$. Then one finds a subset of
$u^{\hat{i}}_I$ such that $u^{\hat{i}}_Iu^{\hat{j}}_Jg^{IJ}=0$. This leads to a set of projected $D^{\hat{i}}_\alpha=u^{\hat{i}}_ID^I_\alpha$ which anti-commute:
$$\{D^{\hat{i}}_\alpha,D^{\hat{j}}_\beta\}=-iu^{\hat{i}}_Ig^{IJ}u^{\hat{j}}_J\gamma^{\mu}_{\alpha\beta}\partial_\mu=0.$$
Similarly, we can use harmonic variables to pick out a set of $\eta$'s which anti-commute. In principle, we can choose any four linearly independent combinations of $\eta$'s to define the on-shell superspace, however we would like to have a book-keeping device to restore the $SO(8)$ invariance of the amplitudes, which requires us to break R-symmetry in a covariant manner such as the coset approach. One can also choose to proceed without manifest R-invariance. Then one would be required to impose R-invariance as an additional constraint on the amplitude, as was done for ABJM~\cite{Bargheer:2010hn}.

The harmonic variables we use will be the coset variables for $\frac{SO(8)}{U(1)^4}$~\cite{Ferrara:1999bv}. One introduces three different harmonics, $v^i_I,\,u^a_A,\,\tilde{u}^{\dot{a}}_{\dot{A}}$, corresponding to the three irreducible representations of $Spin(8)$: $8_v,8_s,8_c$. These are 8$\times$8 orthogonal matrices satisfying
\eq
\sum_{I}v^i_Iv^j_I=\delta^{ij},\quad\sum_Au^a_Au^b_A=\delta^{ab},\quad\sum_{\dot{A}}\tilde{u}^{\dot{a}}_{\dot{A}}\tilde{u}^{\dot{b}}_{\dot{A}}=\delta^{\dot{a}\dot{b}}
\label{metric}
\eqe
and are related via
\eq
u^a_A(\Gamma^i)_{a\dot{a}}\tilde{u}^{\dot{a}}_{\dot{A}}=v^i_I(\Gamma^I)_{A\dot{A}}.
\label{conversion}
\eqe
The harmonic variables can be thought of as vielbeins of the coset, with the indices $I,A,\dot{A}$ transforming under the global SO(8). Furthermore, $i,a,\dot{a}$ can be labeled by charges under the four $U(1)$'s. Different combinations of $U(1)$ charges, denoted as $\pm,(\pm),\{\pm\},[\pm]$, give different representations:
\eqa
\nonumber\begin{array}{ccccccccc}8_v\;(i)\rightarrow: & ++ & -- & (++) & (--) & [+]\{+\} & [-]\{-\} & [+]\{-\} & [-]\{+\} \\
8_s\;(a)\rightarrow : & +(+)[+] & +(+)[-] & -(-)[+] & -(-)[-] & +(-)\{+\} & +(-)\{-\} & -(+)\{+\} & -(+)\{-\} \\
8_c\;(\dot{a})\rightarrow : & +(+)\{+\} & +(+)\{-\} & -(-)\{+\} & -(-)\{-\} & +(-)[+] & +(-)[-] & -(+)[+] & -(+)[-]. \end{array}
\eqae
Note that the first four elements in the $8_{v}$ representation have vectorial charge while the last four have spinorial charge. This is because the last four elements form an $SO(4)=SU(2)\times SU(2)$ subgroup, and therefore are more conveniently written in bispinor form. We refer the reader to the original work for a detailed explanation of the decomposition of the charges~\cite{Ferrara:1999bv}.

We will construct the superspace coordinates using the harmonics in vector representation of $Spin(8)$, $v^i_I$. The anti-commutator algebra of $\eta^{i}=v^i_I\eta^I$ is:
\eqa
\{\eta^{++},\eta^{--}\}=\{\eta^{(++)},\eta^{(--)}\}=\{\eta^{[+]\{+\}},\eta^{[-]\{-\}}\}=-\{\eta^{[+]\{-\}},\eta^{[-]\{+\}}\}=1
\eqae
with all other anti-commutators vanishing. From the above equation, we see that one choice of four anti-commuting $\eta$'s is:
$$\eta^{\hat{i}}=v^{\hat{i}}_I\eta^I,\;\hat{i}=1,2,3,4,\rightarrow\eta^{1}\equiv\eta^{++},\;\eta^{2}\equiv\eta^{(++)},\,\eta^{3}\equiv\eta^{[+]\{+\}},\,\eta^{4}\equiv\eta^{[+]\{-\}}.$$
These will be the four fermionic coordinates of the on-shell superspace. Then the superfield has the following expansion:
\eqa
\nonumber\Phi_a(\eta^1,\eta^2,\eta^3,\eta^4)&=&\phi^{'}_{(0)a}+\eta_{\hat{i}}(\Gamma^{\hat{i}})_{a\dot{a}}\psi_{(1)}^{\dot{a}}+\eta_{\hat{i}}\eta_{\hat{j}}(\Gamma^{\hat{i}\hat{j}})_{ab}\phi^{'b}_{(2)}+\eta_{\hat{i}}\eta_{\hat{j}}\eta_{\hat{k}}(\Gamma^{\hat{i}\hat{j}\hat{k}})_{a\dot{a}}\psi_{(3)}^{\dot{a}}\\
&&+\eta^{\hat{i}}\eta^{\hat{j}}\eta^{\hat{k}}\eta^{\hat{l}}(\Gamma^{\hat{i}\hat{j}\hat{k}\hat{l}})_{ab}\phi_{(4)}^{'b}.
\label{canonicalexpansion}
\eqae
The subscripts in parenthesis indicate at which order in the $\eta$ expansion each component appears. Thus there is 1 $\phi^{'}_{(0)a}$, 4 $\psi_{(1)}^{\dot{a}}$, 6 $\phi^{'b}_{(2)}$, 4 $\psi_{(3)}^{\dot{a}}$, and 1 $\phi_{(4)}^{'b}$. Note that the repeated index $\hat{i}$ is summed from $1$ to $4$.

To make the R-indices of the fields manifest, one simply uses the fact that the U(1) charges of each term in the expansion of the superfield are the same. For example, if we choose the on-shell superfield $\Phi_a$ to have the U(1) charges ${+(+)[+]}$, then up to linear order in the $\eta^{\hat{i}}$ expansion we have:
\eqa
\nonumber\Phi_{+(+)[+]}(\eta^1,\eta^2,\eta^3,\eta^4)&=&\phi^{'}_{(0)+(+)[+]}+\eta_{++}\tilde{u}^{+(-)[-]}_{\dot{A}}\psi_{(1)}^{\dot{A}}+\eta_{(++)}\tilde{u}^{-(+)[-]}_{\dot{A}}\psi_{(1)}^{\dot{A}}+\eta_{[+]\{+\}}\tilde{u}^{-(-)\{+\}}_{\dot{A}}\psi_{(1)}^{\dot{A}}\\
&&+\eta_{[+]\{-\}}\tilde{u}^{-(-)\{-\}}_{\dot{A}}\psi_{(1)}^{\dot{A}}+\cdot\cdot\cdot.
\eqae
where $\cdot\cdot\cdot$ represents the higher order terms. This equation is easily established by counting the charges. In particular, $\phi^{'}_{(0)+(+)[+]}=u^{-(-)[-]}_A\phi^A$, and the four $\psi^{\dot{a}}$'s appearing in eq.(\ref{canonicalexpansion}) are
$$\psi^{\dot{a}}_{(1)}=\left\{\begin{array}{c}\psi^{+(-)[-]}=\tilde{u}^{+(-)[-]}_{\dot{A}}\psi^{\dot{A}} \\ \psi^{-(+)[-]}=\tilde{u}^{-(+)[-]}_{\dot{A}}\psi^{\dot{A}} \\ \psi^{-(-)\{+\}}=\tilde{u}^{-(-)\{+\}}_{\dot{A}}\psi^{\dot{A}} \\ \psi^{-(-)\{-\}}=\tilde{u}^{-(-)\{-\}}_{\dot{A}}\psi^{\dot{A}} \end{array}\right..$$
Note that since the gamma matrices in eq.(\ref{canonicalexpansion}) are Clebsch-Gordan coefficients, in the $U(1)^4$ basis they reduce to 1.

The on-shell superspace for $n$ particles is described by the coordinates $\lambda_m$ and $\eta^{\hat{j}}_m$, where $m=1,\cdot\cdot\cdot,n$. Any superamplitude can be written as a function of these on-shell coordinates. Note, however, that a superamplitude encodes the scattering of all possible component fields, or particle species. To extract the scattering amplitude for a particular set of component fields, one simply integrates away the fermionic superspace coordinates while making the following associations:
\eq
1\sim\phi^{'}_{(0)a},\;\eta_m^{\hat{i}}\sim\psi_{m(1)}^{\dot{a}},\;\eta_m^{\hat{i}}\eta_m^{\hat{j}}\sim\phi^{'b}_{m(2)},\;\eta_m^{\hat{i}}\eta_m^{\hat{j}}\eta_m^{\hat{k}}\sim\psi_{m(4)}^{\dot{a}},\;\eta_m^{\hat{i}}\eta_m^{\hat{j}}\eta_m^{\hat{k}}\eta_m^{\hat{l}}\sim\phi_{m(4)}^{'b}.
\label{death}
\eqe
After doing so, the component fields will still be dressed by the harmonic variables. The final step is to restore the R-indices by integrating away the harmonic variables.

We define the harmonic integration, which removes the harmonic variables, following the prescription of the four-dimensional $\mathcal{N}=2$ case~\cite{Galperin:1980fg}
\eqa
\nonumber&&\int dv\; 1=\int du\; 1=\int d\tilde{u}\; 1\equiv1\\
&&\int dv\; v^{i}_I\cdot\cdot v^{j}_J|_{Traceless}=\int du\; u^{a}_A\cdot\cdot u^{b}_B|_{Traceless}=\int d\tilde{u}\; \tilde{u}^{\dot{a}}_{\dot{A}}\cdot\cdot \tilde{u}^{\dot{b}}_{\dot{B}}|_{Traceless}\equiv0.
\eqae
The notation $|_{Traceless}$ indicates that one isolates the part of the integrands that is traceless with respect to the indices $I,A,\dot{A}$. The definition of these harmonic integrals implies that one is picking out pieces that are R-singlets. The R-singlets come from the repeated use of eq.(\ref{metric}) and eq.(\ref{conversion}) on any function that depends on harmonic variables. Note that the harmonic integration will result in all possible R-index contractions between different legs.

\section{S-matrix for 3D $\mathcal{N}=8$ Superconformal Theories}
Equipped with the on-shell superspace, we now introduce the building blocks for the construction of superconformal amplitudes. First we define the supermomentum delta function:
\eqa
\nonumber\delta^2(Q^1)\delta^2(Q^2)\delta^2(Q^3)\delta^2(Q^4)&=&\delta\left(Q^{1\alpha}\right)\delta\left(Q_{\alpha}^{1}\right)\delta\left(Q^{2\beta}\right)\delta\left(Q_{\beta}^{2}\right)\delta\left(Q^{3\gamma}\right)\delta\left(Q_{\gamma}^{3}\right)\delta\left(Q^{4\rho}\right)\delta\left(Q_{\rho}^{4}\right)
\eqae
where
$$Q^{{\hat{i}}\alpha}=\sum_{i=1}^{n}q_{i}^{{\hat{i}}\alpha},\,\,\,\,\,q_i^{{\hat{i}} \alpha}=\lambda_i^{\alpha}\eta^{\hat{i}}_i$$
and the summation is over all external lines, which are labeled by $i$. The $Q^{{\hat{i}}\alpha}$ should not be confused with the supersymmetry generators defined in eq. (\ref{3dgen}). The Lorentz invariant objects are then:
\begin{itemize}
  \item $\delta^3(P)$ and $\delta^2(Q^1)\delta^2(Q^2)\delta^2(Q^3)\delta^2(Q^4)$
  \item $\lambda_i^\alpha\lambda_{j\alpha}=\langle ij\rangle$
  \item $p_i\cdot p_j=-\frac{1}{2}\langle ij\rangle^2$
  \item $q_l^{\hat{i}\alpha}\lambda_{j\alpha}=q_l^{\hat{i}}|j\rangle=\eta_l^{\hat{i}}\langle lj\rangle$
\end{itemize}
With these building blocks, we can now begin our construction of amplitudes.
\subsection{3-point S-matrix}
Lorentz invariance implies that the three-point amplitude should have the following form:
$$A_3=\delta^3(P)f(p_i,q_j,\lambda_k),$$
where $f$ is some general Lorentz-invariant function of $p,q,\lambda$. We also require that this amplitude is dilatation invariant
$$\mathcal{D}A_3=\sum^3_{i=1}D_iA_3=0.$$
Since the momentum delta function has mass dimension $-3$, we can permute the dilatation operator past it, picking up a factor of $-3$:
\eqa
\nonumber\mathcal{D}\delta^3(P)f(p_i,q_j,\lambda_k)&=&\delta^3(P)\left(\mathcal{D}-3\right)f(p_i,q_j,\lambda_k)\\
&=&\delta^3(P)\left(\sum^3_{i=1}\frac{1}{2}\lambda_i^{\alpha}\frac{\partial}{\partial \lambda_i^{\alpha}}\right)f(p_i,q_j,\lambda_k)+(\frac{3}{2}-3)A_3.
\eqae
The factor $\frac{3}{2}$ is due to the constant in the dilatation operator. In order for $A_3$ to be dilatation invariant, the function $f$ must have weight $\frac{3}{2}$ under the operator $\sum^3_{i=1}\frac{1}{2}\lambda_i^{\alpha}\frac{\partial}{\partial \lambda_i^{\alpha}}$, which essentially counts the mass dimension. Since all the building blocks introduced in the previous section have integer mass dimensions, it is impossible to construct a Lorentz-invariant function $f$ such that $\mathcal{D}A_3=0$. Therefore Lorentz and dilatation invariance rule out the possibility of constructing a superconformal on-shell three-point amplitude. In fact, it's not difficult to see that this is true for any odd-point amplitude, since there will be a fractional constant coming from the dilatation operator that cannot be canceled by any rational function of Lorentz invariants.

It should be noted that three-point amplitudes for massless particles in Minkowski space generally vanish by kinematic constraints since momentum conservation implies vanishing Lorentz invariants. For example, $s_{12}=(p_1+p_2)^2=p_3^2=0$. In more than three dimensions, this can be overcome by working with other space-time signatures. For example, in four dimensions, the kinematic invariants can be written in terms of $SL(2,C)$ spinors as follows:
$$s_{ij}=(\lambda_i)^\alpha(\lambda_j)_\alpha(\bar{\lambda}_i)^{\dot{\alpha}}(\bar{\lambda}_j)_{\dot{\alpha}}=\langle ij\rangle[ij].$$
If one works with split signature, then $(\lambda_i)^\alpha\neq(\bar{\lambda}_i)^{\dot{\alpha}}$. As a result, if $s_{ij}=0$ this only implies that $\langle ij\rangle$ or $[ij]$ vanishes while the other can remain nonzero, allowing one to write down non-zero three-point amplitudes.

In three dimensions, all the kinematic invariants will vanish regardless of the signature since $s_{ij}=-\langle ij\rangle^2$. Nevertheless, the analysis we presented in this section is still useful because it carries over to six dimensions, where non-vanishing Lorentz invariants can be defined for three-point amplitudes.
\subsection{4-point S-matrix}
We now move on to the four-point amplitude. In order to have manifest supersymmetry, we assume the amplitude is proportional to both the momentum and supermomentum delta functions:
$$\mathcal{A}_4=\delta^3(P)\delta^8(Q)f(p_i,q_j,\lambda_k)$$
where $\delta^8(Q)=\delta^2(Q^1)\delta^2(Q^2)\delta^2(Q^3)\delta^2(Q^4)$. Applying the dilatation operator gives
\eq
\mathcal{D}\delta^3(P)\delta^8(Q)f(p_i,q_j,\lambda_k)=\delta^3(P)\left(\mathcal{D}-3+4+2\right)f(p_i,q_j,\lambda_k)
\eqe
where the $+4$ comes from the supermomentum delta function and the $+2$ comes from the constant in the dilatation operator. This implies that the function $f$ should have weight $-3$ under $\mathcal{D}$. Since we cannot introduce any more $q$'s \footnote{To see this, we note that from eq.(\ref{death}), the four-point scalar amplitude (which is a component in the superamplitude) carries $\eta^8$.}, this leads to the result that the 4-point amplitude has the following schematic form:
\eq
\mathcal{A}_4\sim\delta^3(P)\delta^8(Q)\frac{1}{\langle ij\rangle\langle kl\rangle\langle mn\rangle}.
\label{ansatz}
\eqe
In the remainder of this subsection, we show that this general form already has full superconformal symmetry. We will fix the precise form through additional symmetry and consistency conditions in the next subsection.

The above amplitude has manifest super Poincar\'e and dilatation invariance. To prove superconformal invariance, one only needs to show that it is invariant under the fermionic generators involving two derivatives. Invariance under the other generators of the superconformal group then follows from the closure of the algebra. We therefore consider
\eqa
\nonumber &&\sum_i\frac{\partial}{\partial (\lambda_i)^\alpha}\frac{\partial}{\partial (\eta_i)^{1}}\mathcal{A}_4=\sum_i\frac{\partial}{\partial (\lambda_i)^\alpha}\frac{\partial Q^{1\beta}}{\partial (\eta_i)^{1}}\frac{\partial}{\partial Q^{1\beta}}\mathcal{A}_4\\
\nonumber &=&\sum_i\left[(\lambda_i)^\beta\frac{\partial}{\partial (\lambda_i)^\alpha}+\delta^\beta_\alpha\right]\frac{\partial}{\partial Q^{1\beta}}\mathcal{A}_4.\\
\eqae
For simplicity, we denote $\mathcal{D}^\beta\,_\alpha=\sum_i(\mathcal{D}_i)^\beta\,_\alpha=\sum_i(\lambda_i)^\beta\frac{\partial}{\partial (\lambda_i)^\alpha}$. Now we look at how $(\mathcal{D}_i)^\beta\,_\alpha$ acts on each part of the amplitude. First consider $\delta^3(P)$:
\eqa
\nonumber\mathcal{D}^\beta\,_\alpha\delta^6(P)&=&\sum_i(\lambda_i)^\beta\frac{\partial P^{\gamma\delta}}{\partial (\lambda_i)^\alpha}\frac{\partial \delta^3(P)}{\partial P^{\gamma\delta}}= 2P^{\beta\delta}\frac{\partial }{\partial P^{\alpha\delta}}\\
\nonumber&=&-3\delta_\alpha^\beta\delta^3(P)\\
\eqae
where in the last line we used $\int dx\; x\partial_x\delta(x)f(x)=-\int dx \delta(x)f(x)$ and
$$\frac{\partial P^{\beta\delta}}{\partial P^{\alpha\delta}}=\delta^\beta_\alpha\frac{1}{2}\frac{\partial P^\mu}{\partial P^\mu}=\delta^\beta_\alpha\frac{3}{2}.$$
Next, let's look at the term involving the supermomentum delta function:
\eqa
\nonumber&&\mathcal{D}^\beta\,_\alpha\frac{\partial}{\partial Q^{1\beta}}\delta^{2}(Q^1)\delta^2(Q^2)\delta^{2}(Q^3)\delta^2(Q^4)\\
\nonumber&=&\left(\sum_i(\lambda_i)^\beta\frac{\partial}{\partial (\lambda_i)^\alpha}\right)2Q^1_\beta\delta^2(Q^2)\delta^{2}(Q^3)\delta^2(Q^4)\\
\nonumber&=&2\frac{\partial}{\partial Q^{1\alpha}}\delta^{2}(Q^1)\delta^2(Q^2)\delta^{2}(Q^3)\delta^2(Q^4)
\eqae
where we've noted that $2Q^{{\hat{i}}\beta}Q^{\hat{i}}_{\alpha}=\delta^\beta_\alpha Q^{{\hat{i}}\gamma}Q^{\hat{i}}_{\gamma}=\delta^\beta_\alpha\delta^{2}(Q^{\hat{i}})$. Finally, let's look at the action of $(\mathcal{D}_i)^\beta\,_\alpha$ on the kinematic factor. Noting that
\eq
\mathcal{D}^\beta\,_\alpha\langle ij\rangle=\delta^\gamma_\alpha\lambda^\beta_i\lambda_{j\gamma}-\delta^\gamma_\alpha\lambda^\beta_j\lambda_{i\gamma}=\delta^\beta_\alpha\langle ij\rangle,
\eqe
we find
\eq
\mathcal{D}^\beta\,_\alpha\frac{1}{\langle ij\rangle\langle kl\rangle\langle mn\rangle}=-3\delta^\beta_\alpha\frac{1}{\langle ij\rangle\langle kl\rangle\langle mn\rangle}.
\eqe
Putting everything together, we have
$$\sum_i\frac{\partial}{\partial (\lambda_i)^\alpha}\frac{\partial}{\partial (\eta_i)^{1}}\mathcal{A}_4=(4-3+2-3)\frac{\partial}{\partial Q^{1\alpha}}\mathcal{A}_4=0.$$
Therefore $\mathcal{A}_4$ is indeed a superconformal invariant.

One can try to extend the ansatz in eq (\ref{ansatz}) to describe n-point amplitudes with $n=2m$, $m>2$. If we assume that the fermionic contribution comes solely from the supermomentum delta function (which would resemble the MHV amplitudes of $\mathcal{N}=4$ sYM), dilatation invariance would restrict the amplitude to have the following schematic form:
$$\mathcal{A}_n\sim \delta^3(P)\delta^8(Q)\frac{1}{\begin{array}{c}\underbrace{\langle ij\rangle\cdot\cdot\cdot \langle lk\rangle} \\ 1+m \end{array}}.$$
For $n>4$, however, this is not a superconformal invariant since
$$\sum_i\frac{\partial}{\partial (\lambda_i)^\alpha}\frac{\partial}{\partial (\eta_i)^{1}}\mathcal{A}_n=(n-3+2-(1+n/2))\frac{\partial}{\partial Q^{1\alpha}}\mathcal{A}_n\neq0, \;{\rm for}\;n\neq4.$$
We therefore conclude that higher-point superconformal amplitudes require additional $q$'s.

\subsection{Consistency Conditions on the 4-point Amplitude}
To find the correct combination of $s_{ij}$'s for the four-point amplitude in eq (\ref{ansatz}), we will implement various consistency conditions. Depending on the gauge algebra from which the interacting theory is built, amplitudes will be required to satisfy certain symmetry properties or identities. We can then check if our basis of  superconformal amplitudes provides a solution to these constraints. Furthermore, since the three-point amplitude vanishes, this imposes a further constraint through factorization; namely that the four-point tree amplitude should have vanishing residue in all channels.

It is a forgone conclusion at the onset that the theory whose amplitudes we are constructing is not a Yang-Mills theory, since the action of Yang-Mills is not even classically conformal in three dimensions. However it is instructive to show that superconformal invariance leads to the violation of an important property of Yang-Mills amplitudes called the photon-decoupling identity.

In a YM theory, the tree-level scattering amplitude for n gluons can be written as
\eq
\mathcal{A}_{YMn}^{tree}=g^{n-2}\sum_{\sigma\in S_n/Z_n}Tr(T^{a_1}\cdot\cdot\cdot T^{a_n})A^{tree}_{YMn}(1,2,\cdot\cdot\cdot,n)
\label{photon}
\eqe
where the $SU(N)$ generators appearing in the trace correspond to the external gluons and the sum is over all non-cyclic permutations of the external legs. This decomposition of the of amplitude is known as color-ordering and $A^{tree}_{YMn}$ is referred to as the color-ordered partial amplitude ($\mathcal{A}_{YMn}^{tree}$ is called the color-dressed amplitude). Note that the sum can be implemented by keeping the first external leg fixed and summing over all permutations of legs 2 though n. Since the trace is cyclic symmetric, the same is true for the color-ordered amplitude. Also note that $A^{tree}_{YMn}(1,2,\cdot\cdot\cdot,n)$ can be taken to be planar since the non-planar contributions to the color-dressed amplitude will come from the non-cyclic permutations.

If we replace one of the generators by the identity matrix (which corresponds to introducing a $U(1)$ gauge field), then eq.(\ref{photon}) vanishes since the photon decouples. This is the photon-decoupling identity \cite{Bern:1990ux}. Choosing $T^{a1}$ to be $1$, the color trace becomes a trace over the remaining $n-1$ objects, and cyclic rotation of these $n-1$ generators will have identical color factors. In this case, eq.(\ref{photon}) can only vanish if the partial amplitudes with the same color factors add to zero. For a four-point amplitude, this leads to the identity
\eq
A(1,2,3,4)+A(1,3,4,2)+A(1,4,2,3)=0.
\label{decouple}
\eqe
If the three-dimensional theory we are studying in this paper corresponds to a non-abelian YM theory, then we should be able to choose the $s_{ij}$'s in eq (\ref{ansatz}) such that eq.(\ref{decouple}) is satisfied.

If we specialize eq (\ref{ansatz}) to be a color-ordered amplitude, that means that it cannot have poles in the u channel since this would correspond to a non-planar contribution:
\eq
A_4(1,2,3,4)\sim\delta^3(P)\delta^8(Q)\left(\alpha\frac{\langle 12\rangle}{st}+\beta\frac{\langle 13\rangle}{st}+\gamma\frac{\langle 14\rangle}{st}\right).
\label{wrong}
\eqe
Using momentum conservation one can show that
\eq
\langle12\rangle=\langle34\rangle,\;\langle23\rangle=\langle14\rangle,\;\langle13\rangle=\langle42\rangle.
\label{identify}
\eqe
Note that while $\langle12\rangle=\langle34\rangle$ is only true up to a sign, this sign will fix the relative sign of the other spinor inner products through momentum conservation. For example, if $\langle12\rangle=\langle34\rangle$, then we have
\eqa
\langle12\rangle\langle23\rangle=-\langle14\rangle\langle43\rangle=\langle14\rangle\langle12\rangle\rightarrow\langle23\rangle=\langle14\rangle.
\eqae
Plugging eq. (\ref{wrong}) into eq.(\ref{decouple}) and using the relations in eq (\ref{identify}) gives $\alpha=\beta=\gamma$. This leads to
$$A_4(1,2,3,4)\sim\delta^3(P)\delta^8(Q)\left(\frac{\langle 12\rangle+\langle 13\rangle+\langle 14\rangle}{st}\right).$$
Since we are constructing a color-ordered amplitude, it should be cyclic symmetric, but this solution is obviously not. Therefore we find that there is no superconformal solution that satisfies both cyclic symmetry and the photon-decoupling identity. This is consistent with the fact that YM in three dimensions is not classically conformal.

We now consider the constraint coming from factorization. Since we found that the three-point amplitude vanishes, this requires the residue of $\mathcal{A}_4$ to vanish in all channels as well. A simple analysis will show that our ansatz for the color-ordered amplitude in eq.(\ref{wrong}) does not admit a cyclic symmetric solution with the correct factorization property, so we will consider the color-dressed amplitude for which we propose the following ansatz:
$$\mathcal{A}_4\sim\delta^3(P)\delta^8(Q)\frac{f(\lambda_i)}{stu}.$$
Here $f(\lambda_i)$ is a polynomial of spinor inner products with mass dimensions three. The requirement of vanishing residues uniquely fixes $\mathcal{A}_4$ up to a constant:\footnote{We note that there may be an issue of whether or not the residues vanish fast enough, however this is the only possibility which can have vanishing residues. Furthermore, in the next section we will show that this result agrees with the four-point amplitude of the BLG theory, which also has a vanishing three-point amplitude.}
\eq
\mathcal{A}_4\sim\delta^3(P)\delta^8(Q)\frac{\langle12\rangle\langle13\rangle\langle23\rangle}{stu}
\label{superBLG}
\eqe
For example the $s$-channel residue, $r_s\sim\frac{\langle12\rangle\langle13\rangle\langle23\rangle}{tu}$, vanishes since
$$s=0\rightarrow\langle12\rangle=0.$$
The same applies to all other channels.

Notice that eq.(\ref{superBLG}) is totally antisymmetric with respect to the exchange of any pair of indices. Indeed,
\eqa
\nonumber1\leftrightarrow 2: &&\mathcal{A}_4=\delta^3(P)\delta^8(Q)\frac{\langle12\rangle\langle13\rangle\langle23\rangle}{stu}\rightarrow \delta^3(P)\delta^8(Q)\frac{\langle21\rangle\langle23\rangle\langle13\rangle}{stu}=-\mathcal{A}_4\\
\nonumber1\leftrightarrow 4:&& \mathcal{A}_4=\delta^3(P)\delta^8(Q)\frac{\langle12\rangle\langle13\rangle\langle23\rangle}{stu}\rightarrow \delta^3(P)\delta^8(Q)\frac{\langle42\rangle\langle43\rangle\langle23\rangle}{stu}=-\mathcal{A}_4,\\
\eqae
where we used eq.(\ref{identify}) in the last line. This implies that the three-dimensional theory can be formulated in terms of a totally anti-symmetric four-index structure constant $f^{abcd}$. Such a structure
constant also appears in the BLG theory when it is formulated as an $SO(4)$ gauge theory (see Appendix \ref{APPA} for more details). Later, we will demonstrate that our four-point amplitude matches the component result of the BLG model. Our analysis therefore picks out the BLG theory, suggesting that it is the only three-dimensional theory with maximal superconformal symmetry that admits a Lagrangian description.\footnote{Note that the structure constant in the BLG theory is constrained to be $\epsilon^{abcd}$ by the fundamental identity in eq.(\ref{fund}). At the moment, our analysis does not constrain the structure constant to obey the fundamental identity, but we expect that this constraint will appear after constructing the six-point amplitude and imposing various consistency conditions.} From this point of view, it is not surprising that we found that all odd-point amplitudes must vanish. In particular, since all odd-point amplitudes in the BLG theory have a Chern-Simons gauge field in at least one of their external legs, they must vanish since the gauge fields have no propagating degrees of freedom.

\subsection{Comparison to BLG} 
In this subsection, we will compute the tree-level four-point scalar amplitude of BLG using Feynman diagrams and match it with the scalar component of the four-point superamplitude constructed in the previous subsection. It follows from supersymmetry that all the other components in the expansion of the four-point superamplitude match the BLG theory as well.

Before describing the four-point calculation, we would like to point out why the BLG theory evades the photon-decoupling identity described in the previous subsection. This can be seen group theoretically by noting that the BLG theory can be written as an $SO(4)$ gauge theory with matter in fundamental representation. In this formulation, the photon-decoupling identity is not well-defined because one cannot define color-ordering. On the other hand, it is also possible to write the BLG theory as an $SU(2) \times SU(2)$ gauge theory where the matter transforms in the bi-fundamental representation of $SU(2) \times SU(2)$. In this case, color-ordering can be defined (as demonstrated in Appendix \ref{APPB}). Nevertheless, the photon-decoupling identity is evaded. The basic reason is that if one adds a U(1) component to the gauge field in a pure Yang-Mills theory it will completely decouple, but if one adds a component proportional to the unit matrix to one of the scalar fields in the BLG theory it will not decouple. As a result, if we replace one of the generators associated with an external leg in a four-point amplitude by the unit matrix, the amplitude must vanish in a pure Yang-Mills theory but not in the BLG theory. Similar arguments apply for the ABJM theory.

It is easiest to compute the four-point scalar amplitude when the BLG theory is written as an $SO(4)$ gauge theory. For the analogous calculation carried out using bi-fundamental notation, see Appendix \ref{APPB}. The terms in the action that are needed to compute the amplitude are
\[ -\frac{1}{2}\partial_{\mu}X^{Ia}\partial^{\mu}X^{Ia}+\frac{1}{8}\epsilon^{\mu\nu\lambda}\epsilon^{abcd}A_{\mu ab}\partial_{\nu}A_{\lambda
cd}-g\partial_{\mu}X^{Ia}X^{Ib}A_{\mu ab}\]
where $I=1,...,8$ and $a=1,...,4$. The covariant derivative of this theory is given by $D_{\mu}X_{a}^{I}=\partial_{\mu}X_{a}^{I}+gA_{\mu ab}X^{I b}$. Note that $A_{\mu ab}=-A_{\mu ba}$, so the covariant derivative can also be written as $D_{\mu}X_{a}^{I}=\partial_{\mu}X_{a}^{I}+g\epsilon_{abcd}\tilde{A}_{\mu}^{cd}X^{I b}$, which matches the conventions of various other papers. See appendix \ref{APPA} for more details. Since the gauge field only appears on internal lines, its normalization isn't important. One can also add the following gauge-fixing term \cite{Gustavsson:2008bf}:
\[ \mathcal{L}_{gf}=\frac{1}{2\xi}\epsilon^{abcd}\partial_{\mu}A_{ab}^{\mu}\partial_{\nu}A_{cd}^{\nu}+ghosts\]
where $\xi$ is a gauge-fixing parameter. After transforming to momentum space, the kinetic and gauge-fixing terms for the gauge field become
\[ -\frac{1}{2}A_{\mu ab}(p)K^{\mu\nu abcd}A_{\nu cd}(-p)\] where
\[ K^{\mu\nu abcd}=\epsilon^{abcd}\left(-\frac{i}{4}\epsilon^{\mu\nu\lambda}p_{\lambda}+\frac{1}{\xi}p^{\mu}p^{\nu}\right){\normalcolor .}\] The
propagator for the gauge field is therefore\[ -i\left(K^{-1}\right)_{\mu\nu
abcd}=2\epsilon_{abcd}\left(\epsilon_{\mu\nu\lambda}p^{\lambda}-\frac{i\xi}{4}\frac{p_{\mu}p_{\nu}}{p^{2}}\right)\frac{1}{p^{2}}{\normalcolor .}\] Furthermore, the 3-point vertex depicted in Figure 1 is given by\[ -g\left(p_{a}-p_{b}\right)_{\mu}\delta^{IJ}{\normalcolor .}\] Note that all momenta are taken to be outgoing.
\begin{figure}[tb]
\center
\includegraphics [height=25mm]{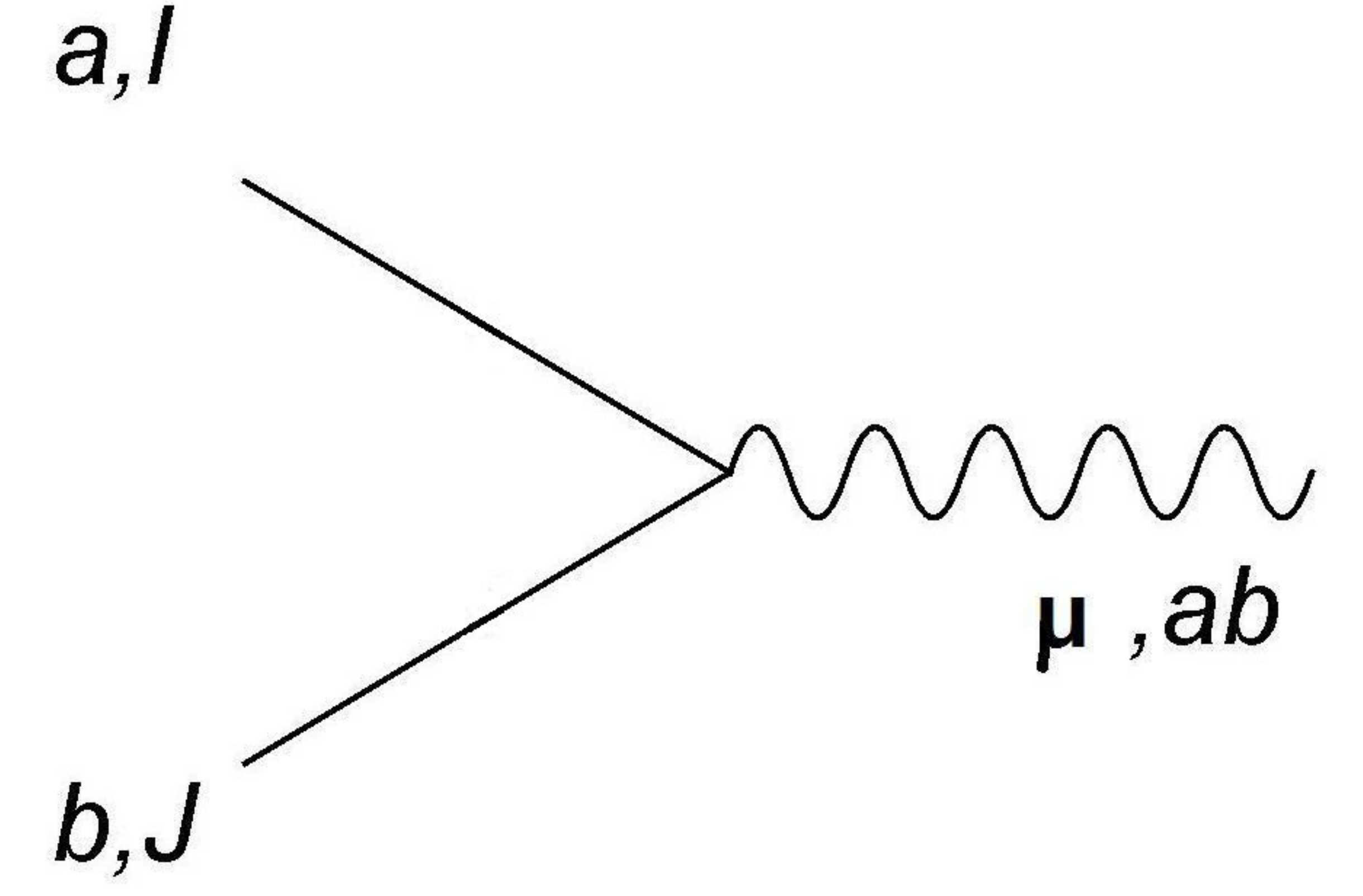}
\caption{3-pt vertex for SO(4) BLG theory} \label{F1}
\end{figure}
\begin{figure}[tb]
\center
\includegraphics [height=30mm]{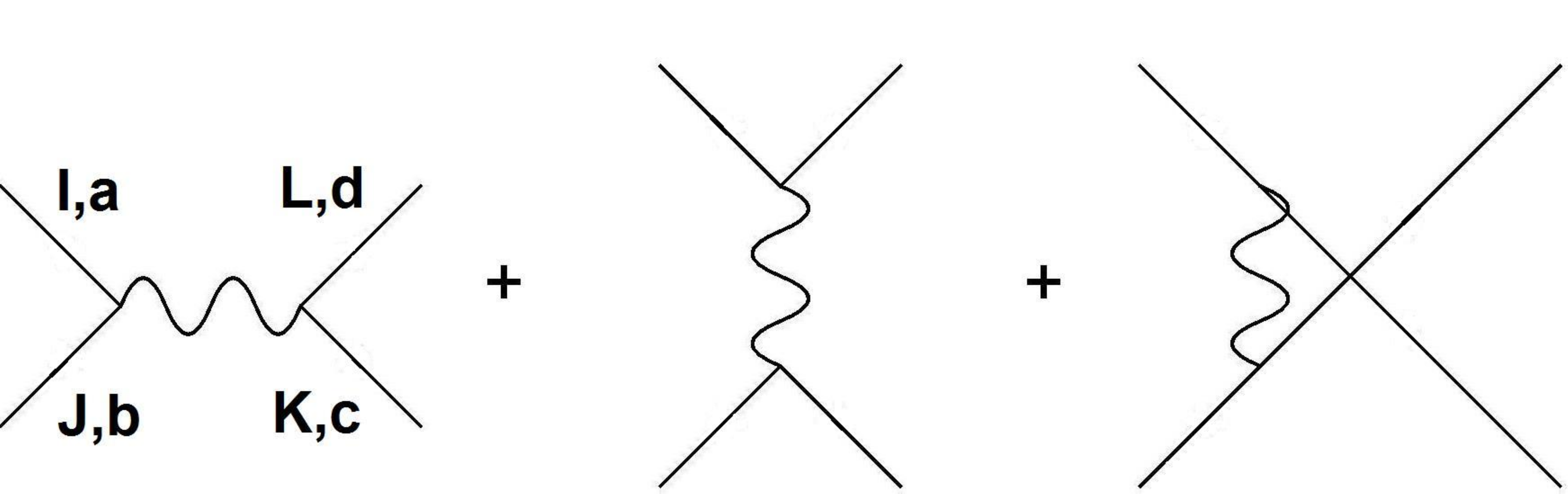}
\caption{Tree-level 4-pt scalar amplitude for SO(4) BLG theory}
\label{F1}
\end{figure}

The tree-level four-point scalar amplitude is given by the sum of three diagrams depicted in Figure 2. Using the Feynman rules described above, one finds that the amplitude is given by
\[\mathcal{A}_{4}=4g^{2}\delta^3(P)\epsilon_{abcd}\epsilon^{\mu\nu\lambda}p_{a\mu}p_{b\nu}\left(p_{c}-p_{d}\right)_{\lambda}\left[\frac{\delta^{IJ}\delta^{KL}}{\left(p_{a}+p_{b}\right)^{2}}+\frac{\delta^{IL}\delta^{KJ}}{\left(p_{a}+p_{d}\right)^{2}}+\frac{\delta^{IK}\delta^{JL}}{\left(p_{a}+p_{c}\right)^{2}}\right].\]
To compare this to the scalar component of the superamplitude, we must sum over all contractions of the R-indices (for reasons that we explain shortly). Doing so gives
\[\mathcal{A}_{4} \sim \delta^3(P)\epsilon_{abcd}\epsilon^{\mu\nu\lambda}p_{a\mu}p_{b\nu}\left(p_{c}-p_{d}\right)_{\lambda}\left[\frac{1}{\left(p_{a}+p_{b}\right)^{2}}+\frac{1}{\left(p_{a}+p_{d}\right)^{2}}+\frac{1}{\left(p_{a}+p_{c}\right)^{2}}\right].\]
Next, we must express this amplitude in twistor space. Noting that $\epsilon^{\mu\nu\lambda}p_{a\mu}p_{b\nu}\left(p_{c}-p_{d}\right)_{\lambda}$ is totally antisymmetric under exchanges of the external legs and recalling that $\langle12\rangle\langle13\rangle\langle23\rangle$ is also totally antisymmetric, it follows that
\begin{equation}
\mathcal{A}_4\sim \delta^3(P) \langle12\rangle\langle13\rangle\langle23\rangle\left(\frac{1}{s}+\frac{1}{t}+\frac{1}{u}\right).
\label{component}
\end{equation}

To extract the scalar component of the superamplitude in eq.(\ref{superBLG}), we must integrate out the fermionic superspace components as well as the harmonic variables. Integrating out the harmonic variables extracts the R-singlet piece of the four-point scalar amplitude, as explained in section 3. As a result, we will be left with a sum over all different ways in which the R-indices of the external legs can be contracted:
$$\mathcal{A}_4|_{\psi\rightarrow0}=A(\phi^I_1\phi_{2I}\phi^J_3\phi_{4J})+A(\phi^I_1\phi_{2J}\phi^J_3\phi_{4I})+A(\phi^I_1\phi^J_{2}\phi_{3I}\phi_{4J}).$$
Recalling the identification in eq.(\ref{death}), this corresponds to integrating out the following combinations of $\eta$'s from the supermomentum delta function appearing in the superamplitude:
\begin{eqnarray}
\nonumber
\left(\int d\eta^1_id\eta^2_id\eta^3_id\eta^4_i+\int d\eta^1_id\eta^2_id\eta^3_jd\eta^4_j+\int d\eta^1_jd\eta^2_id\eta^3_id\eta^4_j+\int d\eta^1_id\eta^2_jd\eta^3_id\eta^4_j\right)\delta^4(Q^\alpha)\\
\nonumber
\times\left(\int d\eta^1_kd\eta^2_kd\eta^3_kd\eta^4_k+\int d\eta^1_kd\eta^2_kd\eta^3_ld\eta^4_l+\int d\eta^1_ld\eta^2_kd\eta^3_kd\eta^4_l+\int d\eta^1_kd\eta^2_ld\eta^3_kd\eta^4_l\right)\delta^4(Q_\alpha)\\
\label{etas}
\end{eqnarray}
where $\delta^4(Q^\alpha)=\delta(Q^{1\beta})\delta(Q^{2\gamma})\delta^4(Q^{3\delta})\delta^4(Q^{4\sigma})$ and we sum over all possible
assignments of $i,j,k,l$. There are a total of $4! = 24$ assignments of the external legs.
For the choice $\{i,j,k,l\}=\{1,2,3,4\}$, the integral in eq. (\ref{etas}) gives $4u^2$.
After taking into account the other possibilities (which simply permute the kinematic variables), we find that the scalar component of the superamplitude in eq.(\ref{superBLG}) is proportional to
\eq
\delta^3(P)\frac{\langle12\rangle\langle13\rangle\langle23\rangle(s^2+t^2+u^2)}{stu}.
\eqe
Noting that $s^2+t^2+u^2=-2(tu+ts+su)$, this matches the result we obtained from the BLG theory in eq. (\ref{component}).

In conclusion, we see that by imposing symmetry and consistency conditions on the S-matrix, we arrive at the four-point amplitude of the BLG theory, which is the only known maximal superconformal theory in three dimensions that admits a Lagrangian description.
\subsection{ABJM Four-point Amplitude from BLG}
Our approach can be easily extended to theories with fewer supersymmetries like the ABJM theory, which has $\mathcal{N}=6$ supersymmetry.
Whereas the on-shell multiplet in a maximal theory can be represented with a single superfield, in theories with less-than-maximal supersymmetry, multiple superfields are required to encode the on-shell multiplet, since it is not self-CPT. Furthermore, if a lower supersymmetric theory can be embedded in a maximal one, then one can obtain the amplitudes of the theory with less susy by truncating the maximal amplitudes according to the particle species. Since the particle species are dictated by the on-shell superspace variables, truncation corresponds to integrating away part of the superspace variables. \footnote{For a detailed discussion of obtaining lower supersymmetric amplitudes from maximal ones, see~\cite{Bern:2009xq, supersum}.} In this section, we will demonstrate that the four-point superamplitude of ABJM (which was constructed in \cite{Bargheer:2010hn}) can be derived from the BLG theory by integrating out one of the fermionic on-shell superspace coordinates.

Since the ABJM theory has $SU(4)$ R-symmetry, we can introduce $\frac{SU(4)}{U(1)^3}$ harmonics to construct its on-shell superspace by analogy with what we did in section 3. One finds that there are three fermionic on-shell superspace coordinates and that the on-shell multiplet can be encoded in two superfields:
\eqa
\nonumber \mathcal{N}=6&\rightarrow&\begin{array}{c}\hat{\Phi}(\eta)=\phi+\psi_I\eta^I+\eta^I\eta^J\phi_{IJ}+\eta^I\eta^J\eta^K\psi_{IJK} \\\quad \hat{\Psi}^{IJK}(\eta)=\bar{\psi}^{IJK}+\eta^I\bar{\phi}^{JK}+\eta^I\eta^J\bar{\psi}^{K}+\eta^I\eta^J\eta^K\bar{\phi}\end{array}
\eqae
where $I=1,2,3$ labels the three independent $\eta$'s. Comparing this with eq.(\ref{canonicalexpansion}), we see that we can identify the two $\mathcal{N}=6$ superfields as a subset of our $\mathcal{N}=8$ superfield by removing one of the four $\eta$'s, which we choose to be $\eta^{++}$. This can be achieved either by setting it to zero or by integrating it away:
\eq
\hat{\Phi}(\eta)=\Phi(\eta)|_{\eta^{++}=0},\;\;\hat{\Psi}^{IJK}=\int d\eta^{++} \Phi(\eta).
\eqe

We can therefore obtain $\mathcal{N}=6$ amplitudes from $\mathcal{N}=8$ amplitudes by integrating away $\eta^{++}$. Since there are two $\eta^{++}$'s coming from supermomentum delta function in the four-point amplitude in eq (\ref{superBLG}), one needs to choose two legs, $\eta_i^{++}$ and $\eta_j^{++}$, to integrate. This corresponds to identifying legs $i$ and $j$ to be in the multiplet carried by $\hat{\Psi}^{IJK}$. Choosing $i=2$, $j=4$ we see that
$$\int d\eta^{++}_2d\eta^{++}_4\mathcal{A}_4=\delta^{3}(P)\delta^{6}(Q)\frac{\langle12\rangle\langle23\rangle\langle13\rangle\langle24\rangle}{stu}=\delta^{3}(P)\delta^{6}(Q)\frac{\langle12\rangle\langle23\rangle}{st}$$
which is indeed the four-point color-ordered superamplitude of ABJM. Hence, if we expand our $\mathcal{N}=8$ amplitude into different sets of $\eta_i^{++}\eta_j^{++}$, the coefficients are simply different color-ordered ABJM amplitudes.\footnote{Note that the structure of the gauge group of ABJM is encoded in the four-index structure constant implied by the four-point superamplitude, which is not totally antisymmetric.} Interestingly, this implies that the Yangian symmetry found in ABJM~\cite{Bargheer:2010hn} is also present in BLG. Note that the Yangian generators in BLG are constructed from the generators of an $OSp(6|4)$ subgroup of the full superconformal symmetry group $OSp(8|4)$. There is a group-theoretic obstruction to extending the Yangian symmetry to $OSp(8|4)$~\cite{Drummond:2009fd}.

Although BLG and ABJM have the same on-shell degrees of freedom, they differ in the structure of the gauge group and the R-symmetry. While the gauge group of the former is $SU(2)\times SU(2)$, the latter can have any $U(N)\times U(N)$ as its gauge group. At tree-level, different choices of $N$ with $N\geq2$ are equivalent, therefore the difference between ABJM and BLG is really in the $U(1)\times U(1)$ part. Furthermore, if we denote the sum and difference of these two $U(1)$ fields by $\hat{A}_\pm$, the kinetic term for the gauge fields takes the form $\hat{A}_- \wedge d\hat{A}_+$ and only $\hat{A}_-$ couples to the matter fields. As a result, the $U(1)\times U(1)$ gauge fields do not contribute to the ABJM scattering amplitudes. The difference in R-symmetry results in the reorganization of external states since, for example, the eight scalars in BLG can be linearly combined into the four complex scalars of ABJM~\cite{Benna:2008zy}. It is therefore not surprising that the BLG four-point amplitude can be expressed as a linear combination of ABJM amplitudes at tree-level.

\section{6D Amplitudes in Supertwistor Space}
In previous sections we've seen how one can glean information about the structure of an action through the study of consistent S-matrix elements. In this section and the next two sections we will apply the same approach to the case of maximal superconformal theories in six dimensions.

First we give a brief introduction to the spinor-helicity formalism in six dimensions. For more details see \cite{Cheung:2009dc, Dennen:2009vk, Boels:2009bv}. In six dimensional Minkowski space, the covering group of the Lorentz group is $SU^*(4)$. Therefore a vector is in the anti-symmetric representation of $SU^*(4)$ and the inner product of two vectors is defined as a contraction with the $SU^{*}(4)$ invariant tensor $\e_{ABCD}$
$$V^\mu\rightarrow \begin{array}{c}\displaystyle{\not\,}V^{AB}=V^{\mu}\Sigma_{\mu}^{AB} \\ \displaystyle{\not\,}V_{AB}=V^{\mu}\tilde{\Sigma}_{\mu AB}\end{array},\;V^\mu U_\mu=\begin{array}{c}-2\e^{ABCD}\displaystyle{\not\,}V_{AB}\displaystyle{\not\,}U_{CD} \\-4\displaystyle{\not\,}V^{AB}\displaystyle{\not\,}U_{AB}\end{array}$$
where the $\Sigma^\mu$'s and $\tilde{\Sigma}^\mu$'s are 4$\times$4 antisymmetric matrices that satisfy the Clifford algebra
$$\Sigma^{\mu}\tilde{\Sigma}^\nu+\Sigma^\nu\tilde{\Sigma}^\mu=2\eta^{\mu\nu}$$
and
$$\Sigma^{\mu AB}\Sigma_{\mu}^{CD}=-2\e^{ABCD},\;\Sigma^{\mu AB}\tilde{\Sigma}_{\mu CD}=-2(\delta^A_C\delta^B_D-\delta^B_C\delta^A_D).$$
For simplicity, we drop the $^*$ from now on keeping in mind that the spinors are pseudoreal.

The six dimensional on-shell massless condition can be solved using chiral spinors. From the above relations we see that
\eq
p^2=-2 \epsilon_{ABCD}\displaystyle{\not\,}p^{AB}\displaystyle{\not\,}p^{CD}=0\rightarrow \displaystyle{\not\,}p^{AB}=\lambda^{A a}\lambda^B\,_{a},\;\;
\eqe
where $a$ and in latter examples $\dot{a}$ are $SU(2)$ indices.\footnote{ One can work in other signatures. For example, in the Wick rotated spacetime with Lorentz group $SO(3,3)$ (whose covering the covering group is $SL(4,R)$), the spinors are real and $a,\dot{a}$ are $SL(2,R)$ indices.} The bi-spinor form of the momentum solves the on-shell constraint since there are no four-component totally anti-symmetric tensors in $SU(2)$. One can also represent the momentum in the anti-fundamental representation:
\eq
\displaystyle{\not\,}p_{AB}=\frac{1}{2}\epsilon_{ABCD}\displaystyle{\not\,}p^{CD}=\tilde{\lambda}_A\,^{\dot{b}}\tilde{\lambda}_{B\dot{b}},\;\; \lambda^A\,_a\tilde{\lambda}_{A\dot{a}}=0.
\eqe
This solution can also be understood by counting components. A null vector in six dimensions has five components including a scale factor, meanwhile $\lambda^{Aa}$ has $4\times2=8$ components and the $SU(2)$ invariance removes three of them. Since the definition of the little group is the set of transformations that leave the null momentum invariant, the $SU(2)$ indices on the spinors correspond to the six dimensional little group $SO(4)=SU(2)\times SU(2)$.

\subsection{6D Superconformal Group}
The supertwistors for a maximal superconformal theory in six dimensions with $Sp(4)$ R-symmetry are in the spinor representation of the supergroup $OSp^*(8|4)$:
$$\zeta^{\mathcal{M}a}=\left(\begin{array}{c}\xi^{\mu a} \\ \eta^{Ia} \end{array}\right),\;\;\mu=1,\cdot\cdot,8, \;I=1,\cdot\cdot,4,\;\; a=1,2.$$
The bosonic part of the twistor $\xi^{\mu a}$  is an eight dimensional chiral spinor transforming under $SO(2,6)=SO^*(8)$, while the fermionic part $\eta^{Ia}$ is a four dimensional spinor transforming under $USp(4)$. The six-dimensional chiral spinors introduced in the previous section can be viewed as half of the $\xi^{\mu a}$'s, which can be decomposed into a pair of six-dimensional chiral and anti-chiral spinors~\cite{Penrose:1986ca}. The additional $SU(2)$ index $a$ comes from the fact that the representation is pseudoreal, so the spinor and it's complex conjugate form a $SU(2)$ doublet. The superconformal generators are represented as
$$G_{\mathcal{M} \mathcal{N}}=\zeta_{[\mathcal{M}}^{\,\,\,\,\,\,\,\,\,\,a}\zeta_{\mathcal{N})a}$$
where one symmetrizes with respect to $Sp(4)$ indices and antisymmetrizes with respect to $SO(8)$ and mixed indices.

The spinors are self-conjugate:
$$[\zeta^{\mathcal{M}a},\zeta^b_{\mathcal{N}}\}=\delta^{\mathcal{M}}\,_{\mathcal{N}}\epsilon^{ab}\rightarrow [\xi^{\mu a},\xi^{\nu b}]=\eta^{\mu\nu}\epsilon^{ab},\;\{\eta^{Ia},\eta^{Jb}\}=\epsilon^{ab}\Omega^{IJ}$$
where $\eta^{\mu\nu},\;\epsilon^{ab},\;\Omega^{IJ}$ are $SO^*(8)$, $SU(2)$, and $USp(4)$ metrics respectively. Note that we can separate the independent degrees of freedom by their weights in the $SU(2)$ little group, i.e. their charges under $J_z$. Then we have
$$[\zeta^{\mathcal{M}+},\zeta^-_{\mathcal{N}}\}=\delta^{\mathcal{M}}\,_{\mathcal{N}}\epsilon^{+-}$$
where $\epsilon^{+-}=-\epsilon^{-+}$, and $\zeta^{\mathcal{M}+}$ can be chosen to carry the independent degrees of freedom.

Decomposing the twistors into six dimensional spinors one has
$$\left(\begin{array}{c}\zeta^{\mu a} \\ \eta^{Ia} \end{array}\right)\rightarrow \left(\begin{array}{c} \lambda^{Aa} \\ \mu^a_{B} \\ \eta^{Ia}  \end{array}\right),\;\;A,B=1,\cdot\cdot 4$$
where $A,B$ are Lorentz indices and $a$ is the chiral part the little group $SO(4)=SU(2)\times SU(2)$. The canonical commutation relation becomes
$$[\lambda^{Aa},\mu^b_{B}]=\delta^A_B\epsilon^{ab}\rightarrow \mu_{Aa}=-\frac{\partial}{\partial \lambda^{Aa}}.$$
We can once again use the $SU(2)$ weights to pick out the independent pieces
$$\zeta^{\mathcal{M}+}=\left(\begin{array}{c} \lambda^{A+} \\ \mu^+_{A} \\ \eta^{I+}  \end{array}\right),\quad\zeta^-_{\mathcal{N}}=\left(\begin{array}{c} \mu^-_{B} \\ \lambda^{-B} \\ \eta^{I-}  \end{array}\right).$$

At this point we can write the superconformal generators as generators acting on functions of $\lambda,\eta$:

\eqa
\nonumber P^{AB}&=&\lambda^{Aa}\lambda_{Ba}\quad(6)\\
\nonumber Q^{A}_I&=&\frac{1}{\sqrt{2}}(\lambda^{Aa}\frac{\partial}{\partial \eta^{Ia}}+\lambda^{Aa}\eta^I_a)\quad(16)\\
\nonumber M^A\,_B&=&\lambda^{Aa}\frac{\partial}{\partial\lambda^{Ba}}-\frac{\delta^A\,_B}{4}\lambda^{Ca}\frac{\partial}{\partial\lambda^{Ca}}\quad(16-1=15)\\
\nonumber D&=&\frac{1}{2}\lambda^{Aa}\frac{\partial}{\partial \lambda^{Aa}}+2\quad(1)\\
\nonumber R_{\,\,\,\, J}^{I}&=&\eta^{Ia}\eta_{Ja}+\delta_{J}^{I}\quad(10)\\
\nonumber S_{A}^I&=&\frac{1}{\sqrt{2}}(\eta^{Ia}\frac{\partial}{\partial \lambda^{Aa}}+\frac{\partial}{\partial \lambda^A_{a}}\frac{\partial}{\partial \eta^a_{I}})\quad(16)\\
\nonumber K_{AB}&=&\frac{\partial}{\partial \lambda^{Aa}}\frac{\partial}{\partial \lambda^{B}_a}\quad(6)\\
\label{theman}
\eqae
where the numbers in the parentheses are the number of components. One can check they indeed generate the superconformal algebra
\eqa
\nonumber \{Q^A_I,Q^B_J\}=P^{AB}\Omega_{IJ}&,&\;\;\{S_A^I,S_B^J\}=K_{AB}\Omega^{IJ}\\
\nonumber[D,P^{AB}]=P^{AB}&,&\;\;[ D, K_{AB} ]=-K_{AB}
\eqae
$$[P^{AB},K_{CD}]=\frac{1}{2}\delta^{[A}_{[D}(\delta^{B]}_{C]} D+2M^{B]}_{C]}).$$
The constant in the dilatation operator can be fixed though the algebra in the last line and simply corresponds to the mass dimension of the scalar.

\subsection{Polarization Tensor}
In the previous subsection we demonstrated that momenta can be expressed in terms of twistors and in the next subsection we will show that the supermultiplet can be encoded in a superfield once we construct the on-shell superspace. In order to demonstrate that the S-matrix of the $\mathcal{N}=(2,0)$ theory can be completely described using supertwistor space, what remains to be shown is that the polarization of the two-form gauge field can be expressed in terms of twistors. Recall that the gauge field $A_{\mu\nu}=-A_{\nu\mu}$ ($\mu,\nu=1,2,\cdot\cdot\cdot 5,6$) has a field strength which satisfies the self-duality relation
$$F^{\rho\nu\pi}=\partial^{[\rho}A^{\nu\pi]}=\frac{1}{6}\epsilon^{\rho\nu\pi\mu\sigma\tau}\partial_{[\mu}A_{\sigma\tau]}.$$

The gauge field has a (linearized) gauge transformation $\delta A_{\mu\nu}=\partial_{[\mu}\Lambda_{\nu]}$ which one can fix using the gauge condition
$\partial^\mu A_{\mu\nu}=0$. We therefore propose the following expression for the polarization of the gauge field:
\begin{equation}
A_{\mu\nu}\rightarrow A^{AB,CD}_{ab}=\frac{\left(\lambda^{[A}_{(a}\kappa^{B]c}\right)\left(\kappa^{[C}_{c}\lambda^{D]}_{b)}\right)}{s_{kp}}
\label{polarization}
\end{equation}
where $\kappa^{C}_{c}$ is the spinor of some reference null vector. Note that this is similar to the polarization vector of the on-shell gauge field in six-dimensional sYM~\cite{Dennen:2009vk}. The gauge condition is satisfied since the twistors are solutions to the Dirac equation
$$\displaystyle{\not\,}p_{AB}\lambda^A=0\rightarrow p^{\mu}A_{\mu\nu}=\displaystyle{\not\,}p_{AB}A^{AB,CD}_{ab}=0.$$
Inserting eq. (\ref{polarization}) into the field strength gives
\eqa
\nonumber F_{\rho\nu\pi}\rightarrow F^{[AB],[CD],[EF]}&=&\epsilon^{BDEF}\lambda^{(A}_a\lambda^{C)}_b-\epsilon^{ADEF}\lambda^{(B}_a\lambda^{C)}_b-\epsilon^{BCEF}\lambda^{(A}_a\lambda^{D)}_b+\epsilon^{ACEF}\lambda^{(B}_a\lambda^{D)}_b\\
\nonumber &-&\epsilon^{AECD}\lambda^{(B}_a\lambda^{F)}_b+\epsilon^{BECD}\lambda^{(A}_a\lambda^{F)}_b+\epsilon^{AFCD}\lambda^{(B}_a\lambda^{E)}_b-\epsilon^{BFCD}\lambda^{(A}_a\lambda^{E)}_b\\
\nonumber &+&\epsilon^{CEAB}\lambda^{(D}_a\lambda^{F)}_b-\epsilon^{DEAB}\lambda^{(C}_a\lambda^{F)}_b-\epsilon^{CFAB}\lambda^{(D}_a\lambda^{E)}_b+\epsilon^{DFAB}\lambda^{(C}_a\lambda^{E)}_b.\\
\eqae

\section{6D $\mathcal{N}=(2,0)$ On-shell Superspace}

Since the fermionic twistors $\eta^{Ia}$ introduced in section 5.1 are self-conjugate, we must remove half of them in order to construct the on-shell superspace. This can be achieved by choosing $\eta^{I+}$ as our fermionic coordinates (which breaks the $SU(2)$ little group symmetry), or by using harmonic variables to project out half of the fermionic twistors (which breaks the R-symmetry). We will discuss both choices, although the latter gives simpler results for reasons that will become apparent shortly.

If we choose $\eta^{I+}$ as the fermionic coordinates, then the $USp(4)$ R-symmetry will be preserved and the physical states will be labeled by their representation in the R-symmetry group and their weight in the little group $SU(2)$. In particular, the gauge field is spin-1 and a singlet in $USp(4)$, the scalars are spin-0 and a 5 of $USp(4)$ (which can be represented by a traceless antisymmetric two-index tensor), and the fermions are spin-$\frac{1}{2}$ and a $\bf{4}$ of $USp(4)$. This leads to a superfield with the following expansion:
\eqa
\nonumber\Phi(\eta^{I+})=A^+_{\mu\nu}+\psi_I^{+}\eta^{+I}+\frac{\eta^{+I}\eta^{+J}}{2}[\phi_{IJ}+\Omega_{IJ}A^0_{\mu\nu}]+\frac{\epsilon_{LIJK}\eta^{-I}\eta^{+J}\eta^{+K}}{3!}\psi^{+L}+(\eta^+)^4A^-_{\mu\nu}
\eqae
where $\phi^{IJ}\Omega_{IJ}=0$ and $\pm,0$ indicate the $SU(2)$ weights, i.e. $\pm=\pm 1$ for $\eta$ and $A_{\mu\nu}$, and $\pm=\pm 1/2$ for $\psi$. A similar superfield expansion was found in \cite{Chern:2009nt}. Since the expansion of this superfield begins with the two-form gauge field, the superfield is an R-singlet and has spin 1, indicating that it is an on-shell tensor field.

This choice of the on-shell superspace breaks the $SU(2)$ little group. As a result, the following Lorentz invariants may contribute to amplitudes:
$$\langle ijkl\rangle\equiv\frac{1}{4!}\epsilon_{ABCD}\lambda_i^{A+}\lambda_j^{B+}\lambda_k^{C+}\lambda_l^{D+},\quad[ ijkl]\equiv\frac{1}{4!}\epsilon^{ABCD}\tilde{\lambda}_{iA+}\tilde{\lambda}_{jB+}\tilde{\lambda}_{kC+}\tilde{\lambda}_{lD+}.$$
It would be advantageous to avoid these objects, since their behavior is obscure in collinear limits.
On the other hand, these objects will be absent if we use harmonic variables to construct the on-shell superspace, since this preserves the $SU(2)$ little group. Furthermore, the Lorentz-invariant objects that appear will be easier to analyze in collinear limits, as we demonstrate in section 7.2. We will therefore use harmonic variables to project out half of the fermionic supertwistor components, which will then be used as the fermionic components of the on-shell superspace. After doing so, we will find that the on-shell superfield is a scalar. This is analogous to $\mathcal{N}=4$ sYM, where the on-shell superfield is a scalar in projective superspace and a vector in supertwistor space. As a result, the four-point amplitude of $\mathcal{N}=4$ sYM is simpler in projective superspace, because in supertwistor space one needs additional twistors to carry the spin degrees of freedom \cite{Hatsuda:2008pm}.

The harmonic superspace for off-shell $\mathcal{N}=(2,0)$ supersymmetry was studied in \cite{Ferrara:2000xg}. For the case at hand, the harmonic variables will parameterize the coset $\frac{USp(4)}{U(1)\times U(1)}$. These variables take the form $u^{i}_I$, where the $i$ labels different combinations of $U(1)$ charges and $I$ is the R-index:
$$u^{1}_I=(+,0),\;u^{2}_I=(0,+),\;u^{3}_I=(0,-),\;u^{4}_I=(-,0).$$
They satisfy the following relation:
$$u^i_I\Omega^{IJ}u^j_J=\Omega^{ij}$$
where $\Omega^{ij}$ is a $4\times 4$ matrix of the form
\eq
\Omega^{ij}=\left(\begin{array}{cc}0 & I \\-I & 0\end{array}\right).
\label{harmonics}
\eqe
We can use these harmonic variables to pick out the independent fermionic twistors. If we define
$$\eta^1_a=u^1_I\eta^I_a,\;\;\eta^2_a=u^2_I\eta^I_a,$$
we find that these variables are indeed independent
\eqa
\nonumber&&\{\eta^I_a,\eta^J_b\}=\Omega^{IJ}\epsilon_{ab}\\
&\rightarrow& \{\eta^1_a,\eta^1_b\}=\{\eta^2_a,\eta^2_b\}=\{\eta^1_a,\eta^2_b\}=0
\eqae
since $\Omega^{12}=0$. We will therefore use these twistors as the fermionic coordinates for our on-shell superspace. The bosonic coordinates consist of the $\lambda$'s introduced in section 5.1. In the space of $(\lambda,\eta^1,\eta^2)$, the bosonic generators in eq. (\ref{theman}) remain the same while the fermionic ones become
\eqa
\nonumber Q^{{\hat{i}}A}&=&\lambda^{Aa}\eta^{{\hat{i}}}_a,\quad Q^{A}_{{\hat{i}}}=\lambda^{Aa}\frac{\partial}{\partial\eta^{{\hat{i}}a}}\\
\nonumber S_{A}^{{\hat{i}}}&=&\eta^{{\hat{i}}a}\frac{\partial}{\partial \lambda^{Aa}},\quad S_{{\hat{i}}A}=\frac{\partial}{\partial \lambda^A_{a}}\frac{\partial}{\partial \eta^{{\hat{i}}a}}
\eqae
where ${\hat{i}}=1,2$.
Since the selected $\eta$'s both carry positive $U(1)$ charges and any amplitude can be written as a function of these on-shell coordinates, we will find that amplitudes can be conveniently categorized by their $U(1)$ charges.

Using harmonic variables, the on-shell superfield has the following expansion:
\eqa
\nonumber\Phi(\eta_1,\eta_2)&=&\phi+\psi_1^a\eta^1_{a}+\psi_2^b\eta^2_{b}+[(\eta^1)^2\phi'+(\eta^2)^2\phi''+(\eta^1\cdot\eta^2)\phi'''+\eta^{1a}\cdot\eta^{b2}A_{(ab)}]\\
&&+(\eta^1)^2\eta^{2a}\tilde{\psi}_{2a}+(\eta^2)^2\eta^{a1}\tilde{\psi}_{1a}+(\eta^2)^2(\eta^1)^2\phi''''.
\eqae
In order to restore the R-index structure, one can observe that since the antisymmetric gauge field carries no R-indices, it carries no harmonic variables. This means that the superfield carries $U(1)$ charges (+,+). From this information we deduce that
\eqa
\nonumber \phi&=&u^1_Iu^2_J\phi^{IJ}\\
\nonumber \psi_1^a&=&u^2_I\psi^{Ia},\,\psi_2^a=u^1_J\psi^{Ja}\\
\nonumber \phi'&=&u^4_Iu^2_J\phi^{IJ},\;\phi''=u^3_Iu^1_J\phi^{IJ},\;\phi'''=(u^2_Iu^3_J+u^1_Iu^4_J)\phi^{IJ}\\
\nonumber \tilde{\psi}_{1a}&=&u^3_I\tilde{\psi}^I_{a},\;\tilde{\psi}_{2a}=u^4_I\tilde{\psi}^I_{a}\\
\phi''''&=&u^3_Iu^4_J\phi^{IJ}.
\eqae
With these identifications we can translate between superspace components and the usual components with R-indices.

\section{S-matrix for the 6D $\mathcal{N}=(2,0)$ Theory}
Now that we've shown the that on-shell degrees of freedom and the generators of the superconformal group can be conveniently written in supertwistor space, we can begin our construction of the S-matrix by identifying the fundamental building blocks.

Lorentz and little group invariance requires all $SU(4)$ and $SU(2)$ indices to be contracted.
For amplitudes with more than three external legs, this requires a supermomentum delta function. The case with three external legs is subtle since new variables may arise, as we explain in the next subsection.  The supermomentum delta function takes the form
\eqa
\nonumber\delta^4(Q^1)\delta^4(Q^2)&=&\epsilon_{ABCD}\delta\left(Q^{1A}\right)\delta\left(Q^{1B}\right)\delta\left(Q^{1C}\right)\delta\left(Q^{1D}\right)\epsilon_{EFGH}\delta\left(Q^{2E}\right)\delta\left(Q^{2F}\right)\delta\left(Q^{2G}\right)\delta\left(Q^{2H}\right)
\eqae
where
$$Q^{\hat{i}A}=\sum_{i=1}^{n}q_{i}^{\hat{i}A},\,\,\,\,\,q^{\hat{i}A}_i=\lambda^{Aa}_i\eta^{\hat{i}}_{a},$$
${\hat{i}}=1,2$, and the sum is over all external lines. The other Lorentz invariant objects are:
\begin{itemize}
  \item $\delta^6(P)=\delta^6\left(\sum p\right)$
  \item $s_{ij}=2p_i\cdot p_j$
  \item $q^A(p_i)_{AB}(p_j)^{BC}\cdot\cdot (p_k)_{EF}q^F=q\displaystyle{\not}p_i\displaystyle{\not}p_j\cdot\cdot\cdot\displaystyle{\not}p_kq$
\end{itemize}
Note that there can only be an odd number of momenta between $q$'s because we can only use $\epsilon_{ABCD}$ and $\epsilon^{ABCD}$ to raise and lower $SU(4)$ indices. Furthermore, we do not have spinor inner products in six dimensions, i.e. there is no analog of the three dimensional inner product $\langle ij\rangle$.

With these objects in hand, we may proceed to construct S-matrix elements. Since each on-shell superfield carries charge (+1,+1), an n-point S-matrix element should carry charge (+n,+n). Since the $\eta$'s are the only objects in our twistor space that carry these charges, the $n$-point S-matrix should be proportional to $(q^1)^n(q^2)^n$ (although there are other possibilities when $n=3$). The momentum delta function has mass dimension $-6$ and the constant in the dilatation generator contributes a factor of $2n$ for an $n$-point amplitude. The remaining objects in the amplitude must therefore cancel the $2n-6$ dilatation ``charge" accumulated so far. We begin with the three-point amplitude.
\subsection{3-point S-matrix}
In six dimensions, the vanishing of Lorentz invariants actually introduces new variables, as demonstrated in \cite{Cheung:2009dc}. These variables
(which are only present for the three point amplitudes) may serve as additional building blocks. To see this, we write
$$s_{ij}=(\lambda_i)^{Aa}(\lambda_i)^B_{a}(\tilde{\lambda}_j)^{\dot{b}}_A(\tilde{\lambda}_j)_{B\dot{b}}=\langle i^a|j^{\dot{b}}]\langle i_a|j_{\dot{b}}]=det\langle i|j]=0$$
where we've used $\langle\;|$ to represent $\lambda$ and $[\;|$ for $\tilde{\lambda}$, and the determinant is with respect to the $SU(2)$
indices. This implies that the 2$\times$2 matrix $\langle i_a|j_b] $ is of rank 1, and can be expressed in terms of $SU(2)$ spinors:
\eq
(\lambda_i)^{A}_a(\tilde{\lambda}_j)_{A\dot{a}}= \langle i_a|j_{\dot{a}}] =(u_i)_a(\tilde{u}_j)_{\dot{a}}.
\label{introu}
\eqe
The complete definition and some useful properties of these variables are given in Appendix \ref{APPC}. These variables were used to construct the three-point
amplitude for pure Yang-Mills theory in six dimensions \cite{Cheung:2009dc} as well as the $\mathcal{N}=(1,1)$ theory \cite{Dennen:2009vk}.

We can try to use the $SU(2)$ spinors to construct a three-point amplitude. Since the amplitude carries $U(1)$ charges $(+3,+3)$,
it must be proportional to
$$A_3\sim\delta^6(P)\delta^A(Q^1)\delta^{B}(Q^1)\delta^C(Q^2)\delta^{D}(Q^2)(u_i^a\eta^1_a)(u_j^b\eta^2_b)$$
Noting that the $u$'s have mass dimension $\frac{1}{2}$, the counting for the dilatation operator is $-6+2+1+6=3$ which means that we need a
mass dimension three object in the denominator. Since there is no Lorentz-invariant object of odd mass dimension,
we see that a superconformal three-point amplitude can't be constructed using $SU(2)$ spinors.

Another possibility is to construct the three-point amplitude using 3 $q^{1}$'s and 3 $q^{2}$'s.
Since each $q$ has one free index, there are total of six free indices that need to be contracted. Furthermore, since the only objects we have
to contract these indices are $\epsilon_{ABCD}$, $\epsilon^{ABCD}$, and the momenta (which carry two $SU(4)$ indices), it follows that the amplitude
must contain a factor of the form $q^{\hat{i}}_k\displaystyle{\not}p_lq^{\hat{j}}_m$. There are three possibilities:
\begin{itemize}
  \item If $k \neq m$, this factor will vanish by the Dirac equation. For example,
\eqa
\nonumber q^{\hat{i}}_1\displaystyle{\not}p_2q^{\hat{j}}_3&=&(q^{\hat{i}}_1)^A(\displaystyle{\not}p_2)_{AB}(q^{\hat{j}}_3)^B\\
\nonumber&=&-(q^{\hat{i}}_1)^A(\displaystyle{\not}p_1)_{AB}(q^{\hat{j}}_3)^B-(q^{\hat{i}}_1)^A(\displaystyle{\not}p_3)_{AB}(q^{\hat{j}}_3)^B=0
\eqae
since $(\lambda_i)^{Aa}({\not}p_i)_{AB}=0$.
  \item If $k=m$ and $\hat{i}=\hat{j}$, then
\eq
q^{\hat{i}}_{k}\not p_{l}q^{\hat{i}}_{k}=\eta^{\hat{i}a}_{k}\eta_{k}^{\hat{i}b}\lambda_{ka}^{A}\lambda_{kb}^{B}p_{lAB}=\frac{1}{16}\eta_{k}^{\hat{i}c}\eta_{kc}^{\hat{i}}s_{kl}.
\label{three1}
\eqe
where we used $\eta_{k}^{\hat{i}a}\eta_{k}^{\hat{i}b}=\frac{1}{2}\epsilon^{ab}\eta_{k}^{\hat{i}c}\eta_{kc}^{\hat{i}}$. This factor must vanish because
because $s_{kl}=0$ for a three-point amplitude.
  \item If $k=m$ and $\hat{i} \neq \hat{j}$, then using the $SU(2)$ spinors $u,\tilde{u}$ we find:
\eq
q^{\hat{i}}_{k}\not p_{l}q^{\hat{j}}_{k}|_{s_{lk}=0}=\eta^{\hat{i}}_{ka}u_k^a\tilde{u}_l^{\dot{b}}\tilde{u}_{l\dot{b}}u_k^c\eta^{\hat{j}}_{kc}=0
\label{three2}
\eqe
where in the last equality we used the fact that $\tilde{u}^{[\dot{a}}_l\tilde{u}^{\dot{b}]}_l=0$.
\end{itemize}
We conclude that a rational superconformal three-point amplitude cannot be constructed. Note that this result is independent of the signature or
complexity of spacetime and is basically due to the lack of spinor inner products in the chiral theory.

\subsection{S-matrix for $n>3$}
The fact that the three-point S-matrix vanishes implies that any four-point amplitude must have vanishing collinear limits, i.e. it's residues must vanish in each channel. In this section we will prove that this constraint along with the constraints of rationality and dilatation invariance imply that it is not possible to construct a four-point amplitude. We then generalize this argument to show that all tree-level amplitudes must vanish.

The requirement that a four-point amplitude has (+4,+4) $U(1)$ charges and Poincar\'{e} invariance implies that it consists of
$\delta^6(P)$, 8 $q$'s, and some Lorentz-invariant function of the momenta $f(p)$. From the constraint of dilatation invariance, one finds that the function $f(p)$ should have mass dimension -6. If $f(p)$ is assumed to be a rational function, we can take it to be a fraction whose denominator is a polynomial of the kinematic invariants $s$, $t$, and $u$. Since there are 8 $q$'s in the numerator of the amplitude (each of which has one free index), and the only objects we have to contract these indices are $\epsilon^{ABCD}$, $\epsilon_{ABCD}$, and the momenta in the numerator of $f(p)$ (each of which has two free indices), there are three possibilities:

\begin{itemize}
  \item All eight $q$'s are contracted by two $\epsilon$'s and there are no free indices to be contracted by the momenta. In this case, dilatation invariance implies that $f(p)$ has the schematic form
 $$\delta^6(P)\delta^4(Q^1)\delta^4(Q^2)\frac{1}{s_{ij}s_{kl}s_{mn}}.$$
  \item  Four of the $q$'s are contracted by an $\epsilon$ and the other four are contracted by the momenta in the numerator of $f(p)$. Since there are four free indices to be contracted, there must be at least two momenta in the numerator of $f(p)$. In the minimal case, dilatation invariance implies that the denominator has four kinematic invariants. Since there are only three kinematic invariants available, that means that at least one of them is squared giving a double pole. The schematic form for this case is:
 \eq
 \delta^6(P)\epsilon_{ABCD}\delta^A(Q^1)\delta^B(Q^1)\delta^C(Q^2)\delta^D(Q^2)\frac{(q\displaystyle{\not}pq)(q\displaystyle{\not}pq)}{s_{ij}s_{kl}s^2_{mn}}.
 \label{try1}
 \eqe
  \item All eight $q$'s are contracted by the momenta in the numerator. Since there are eight free indices to be contracted, there must be at least four momenta in the numerator of $f(p)$. In the minimal case, dilatation invariance implies that the denominator has five kinematic invariants. Since there are only three kinematic invariants to work with, at least two of them must be squared, so the amplitude should have the form
\eq
\delta^6(P)\frac{\left[\delta^A(Q)\displaystyle{\not}p_{AB}\delta^B(Q)\right]\left[\delta^C(Q)\displaystyle{\not}p_{CD}\delta^D(Q)\right](q\displaystyle{\not}pq)(q\displaystyle{\not}pq)}{s_{ij}s^2_{kl}s^2_{mn}}.
\label{try2}
\eqe
\end{itemize}
In all three cases, we need at least four $\delta(Q^{A\hat{i}})=\sum_l q^{A\hat{i}}_l$ in order to have manifest invariance under supersymmetry.\footnote{One might think that we need the full $\delta^4(Q^1)\delta^4(Q^2)$ for manifest supersymmetry. However, since each fermionic variable $\eta^a$ has two components from the $SU(2)$ index, $\delta^2(Q)$ is sufficient to localize the $\eta$'s.
Indeed, the three-point amplitude for $\mathcal{N}=(1,1)$ super Yang-Mills in six dimensions only has $\delta^2(Q^1)\delta^2(\tilde{Q})$, where $Q,\tilde{Q}$ reflect the $(1,1)$ supersymmetry~\cite{Dennen:2009vk}. }
We will now demonstrate that the residue constraint cannot be satisfied in any of these cases.

\begin{figure}
\begin{center}
\includegraphics[scale=0.9]{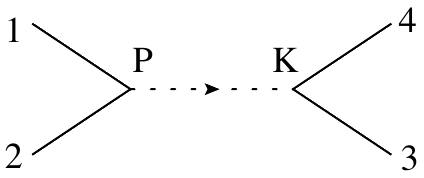}
\caption{Four-point amplitude in the 6d maximal theory. In the limit $s_{12}\rightarrow 0$, we are probing the s-channel residue.}
\label{schannel}
\end{center}
\end{figure}

First note that a rational function of $s_{ij}$'s cannot have vanishing residues.\footnote{For example, $s=0\rightarrow t=-u$. Then any vanishing residue has to be proportional to $t+u$ which is just $s$ and cancels the pole.}
Thus, if $s_{kl}$ appears in the denominator, the numerator must have a term of the form
$q^{\hat{i}}_{l}\left({\not p_{k}}\right)q^{\hat{j}}_{l}$ in order to give a vanishing residue. This is due to the fact
that $q^{\hat{i}}_{l}\left({\not p_{k}}\right)q^{\hat{j}}_{l}$ vanishes when $s_{kl}\rightarrow0$ as shown in eq.(\ref{three1})
and eq.(\ref{three2}). On the other hand, $q^{\hat{i}}_{l}\left({\not p_{k}}\right)q^{\hat{j}}_{m}$ with $l\neq m$ will not vanish when
either $s_{lk}$ or $s_{km}$ goes to zero.
To see this, consider the four-point amplitude depicted in fig.(\ref{schannel}).
In the limit $s_{12}\rightarrow0$, the three momenta at each vertex obey on-shell three-point kinematics.
As a result, we can re-write the object $q^{\hat{i}}_{1}\left({\not p_{2}}\right)q^{\hat{j}}_{3}$ using the $SU(2)$ spinors $u,\tilde{u}$ introduced
in section 7.1 (and described in Appendix C):
$$q^{\hat{i}}_{1}\left({\not p_{2}}\right)q^{\hat{j}}_{3}|_{s_{12}=0}=\eta^{\hat{i}}_{1a}\langle1^a|2^{\dot{a}}][2_{\dot{a}}| 3^b\rangle\eta^{\hat{j}}_{3b}|_{s_{12}=0}=\eta^{\hat{i}}_{1a}u_{1}^a\tilde{u}_2^{\dot{a}}[2_{\dot{a}}| 3^b\rangle\eta^{\hat{j}}_{3b}.$$
Note that $[2_{\dot{a}}| 3^b\rangle$ can't be written in terms of $SU(2)$ spinors because $p_2$ and $p_3$ don't lie on the same vertex.
On the other hand, if we use eq.(\ref{law1}) and note that $\tilde{\lambda}_K=i\tilde{\lambda}_P$ (since $p_K=-p_P$), this allows us to
write everything in terms of $SU(2)$ spinors:
\eqa
\nonumber q^{\hat{i}}_{1}\left({\not p_{2}}\right)q^{\hat{j}}_{3}|_{s_{12}=0}&=&\eta^{\hat{i}}_{1a}u_{1}^a\tilde{u}_P^{\dot{a}}[P_{\dot{a}}| 3^b\rangle\eta^{\hat{j}}_{3b}\\
&=&-i\eta^{\hat{i}}_{1a}u_{1}^a\tilde{u}_P^{\dot{a}}[K_{\dot{a}}| 3^b\rangle\eta^{\hat{j}}_{3b}\\
&=&-i\eta^{\hat{i}}_{1a}u_{1}^a\tilde{u}_P^{\dot{a}}\tilde{u}_{K\dot{a}}u_3^b\eta^{\hat{j}}_{3b}.
\eqae
Furthermore, in Appendix \ref{APPC} we show that $|\tilde{u}_P^{\dot{a}}\tilde{u}_{K\dot{a}}|=\sqrt{-s_{14}}$. Hence
$$s_{12}=0\rightarrow q^{\hat{i}}_{1}\left({\not p_{2}}\right)q^{\hat{j}}_{3}\sim\sqrt{-s_{14}}$$
and similarly
$$s_{23}=0\rightarrow q^{\hat{i}}_{1}\left({\not p_{2}}\right)q^{\hat{j}}_{3}\sim\sqrt{-s_{12}}.$$
Therefore $q^{\hat{i}}_{1}\left({\not p_{2}}\right)q^{\hat{j}}_{3}$ does not vanish on either the $s$ or $t$ channel poles.

From the above discussion, we see that for every factor of the form $s_{lk}$ which appears in the denominator of the amplitude,
the numerator must have at
least one factor of the form $q^{\hat{i}}_{k}\left({\not p_{l}}\right)q^{\hat{j}}_{k}$ in order to have vanishing residues in each channel.
Since this property isn't satisfied by any of the three cases listed above
\footnote{Although some of terms in $\left[\delta^A(Q)\displaystyle{\not}p_{AB}\delta^B(Q)\right]$ will vanish in certain collinear limits, it is
not possible for all of them to vanish simultaneously.  },
we conclude that there is no consistent rational four-point amplitude for the superconformal $\mathcal{N}=(2,0)$ theory.

This analysis can be easily extended to higher-point amplitudes. First of all, an $n$-point amplitude requires $2n$ $q$'s, at least four of
which should appear in a delta function. It follows that one can have a maximum $n-2$ factors of the form
$q^{\hat{i}}_{k}\left({\not p_{l}}\right)q^{\hat{j}}_{k}$ in the numerator of the amplitude. Dilatation invariance then implies that we need at least $2n-4$
factors of the form $s_{kl}$ in the denominator of the amplitude.
The vanishing of all collinear limits then requires that $n-2\geq2n-4\rightarrow n\leq2$. Therefore rationality, dilatation invariance, and
vanishing collinear limits imply that all tree-level amplitudes must vanish. Since rational S-matrix elements probe the structure of the Lagrangian,
this leads us to conjecture that the six-dimensional theory with $OSp^{*}(8|4)$ symmetry does not have a Lagrangian description (at least if one only uses (2,0) tensor multiplets).

\section{Discussion}
In this paper we've established the necessary building blocks for analyzing amplitudes of maximally superconformal theories in
three and six dimensions. In both cases, we find that superconformal invariance implies that the three-point amplitude must vanish.
This behavior is genuine and does not depend on the signature or the complexity of spacetime. The vanishing of the three-point amplitude
puts an additional constraint on the four-point amplitude, notably that it should have a vanishing residue in all channels. In three dimensions,
we find a unique solution for the four-point superamplitude and verify that it agrees with the component result of the BLG model. That the BLG model is selected by our analysis provides strong evidence that it is the only three-dimensional theory with classical $OSp(8|4)$ symmetry that has a
Lagrangian description.

As described in Appendix \ref{APPA}, the BLG theory can be written in terms of a totally anti-symmetric four-index
structure constant which obeys a generalization of the Jacobi-identity called the fundamental identity (see eq. (\ref{fund})). Although the four-point
superamplitude that we constructed implies a totally anti-symmetric four-index structure constant, our analysis does not constrain it to obey the
fundamental identity. In four-dimensional Yang-Mills theory, the Jacobi identity can be derived as a consistency condition of the
BCFW recursion relations~\cite{ Benincasa:2007xk}. As we will see shortly, however, the BCFW approach does not work straight-forwardly in three dimensions.
We anticipate that after deriving the six-point amplitude, the fundamental identity will arise as a consistency condition for the
amplitude to have the correct factorization in all three-particle channels.

In six dimensions, we are unable to find a rational amplitude that is both superconformal and has the correct residues. We therefore conjecture that an interacting six-dimensional Lagrangian with classical $OSp(8|4)$ symmetry cannot be constructed using only $(2,0)$ tensor multiplets, even if the Lagrangian is non-local. A Lagrangian description may exist, however, if one includes additional degrees of freedom. One possibility is that these should be closed strings which couple to the two-form gauge fields of the tensor multiplets, in which case the six dimensional theory is a self-dual string theory \cite{selfdualstring}. It may also be possible to construct a six-dimensional theory with classical $OSp(8|2)$ symmetry which becomes enhanced to $OSp(8|4)$ symmetry at strong coupling using $\mathcal{N}=(1,0)$ tensor multiplets, similar to the ABJM construction in three-dimensions. It is unclear, however, what the gravity dual of such a theory would be. Note that the gravity dual of stack of M2 or M5-branes is M-theory on $AdS_4 \times S^7$ or $AdS_7 \times S^4$, respectively. In the ABJM construction, 1/4 of the supersymmetry is broken on the gravity side by introducing a $\mathbb{Z}_k$ orbifold on the seven-sphere and the supersymmetry only becomes maximal for $k=1,2$, which corresponds to strong-coupling on the field theory side. On the other hand, it is not clear how to implement such an orbifold in $AdS_7 \times S^4$.

Since the three-dimensional amplitudes are superconformal, one might expect that there is a representation in supertwistor space that is similar to the link representation proposed for $\mathcal{N}=4$ sYM \cite{Mason:2009sa, ArkaniHamed:2009dn}. Unfortunately, one of the most useful tools in the analysis of four dimensional amplitudes, the BCFW recursion relations \cite{Britto:2004ap}, cannot be straightforwardly implemented in three dimensions. To see this, note that the BCFW approach requires shifting the momenta of at least two legs into the complex plane while keeping them on-shell. In other words, we must perform the shifts $p_i \rightarrow p_i+zq,\,p_j \rightarrow p_j-zq$ while imposing the constraints $$p_i\cdot q=p_j\cdot q=q^2=0.$$ In three dimensions these constraints imply that $q=0$. Even without BCFW to iteratively generate higher-point tree amplitudes, one can still use the four-point tree-level amplitude to study loop behavior using the generalized unitarity method \cite{Bern:1994zx, Bern:1994cg, Bern:1997sc}. An interesting question is how unitarity methods for loop amplitudes differentiate between the ABJM and BLG theories, since their tree level amplitudes are closely related. Presumably the difference will come from the difference in R-symmetry, which manifests itself through the organization of the on-shell states summed across the cuts.

Although we mainly considered maximal superconformal theories in this paper, our approach can be easily extended to non-maximal
ones like the ABJM theory. In particular, we demonstrated that the tree-level four-point amplitude of the ABJM theory can be obtained from
our $\mathcal{N}=8$ result by integrating out one of the four fermionic coordinates in the on-shell superspace. One can also use
our techniques to analyze the possibility of constructing a six dimensional interacting superconformal action from $\mathcal{N}=(1,0)$ tensor
multiplets \cite{Dall'Agata:1997db}. This theory might be useful for describing phase transitions in six-dimensional compactifications of
string theory and M-theory \cite{1string}. It may also be useful for describing M5-branes, for reasons explained above. Our methods can also be used to analyze amplitudes of non-conformal massless theories such as the world-volume theory for a single M5-brane \cite{m5,Ho:2008ve}.

\section{Acknowledgements}
The work of YH is supported by the US DOE grant DE-FG03-91ER40662 and the work of AEL is supported in part by the US DOE grant DE-FG02-92ER40701. We thank Zvi Bern, Tristan Deenen, Eric D'Hoker and John H. Schwarz for many detailed and enlightening discussions. We would also like to thank Till Bargheer, Pei Ming Ho, Harold Ita, F. Loebbert, Carlo Meneghelli, Donal O'Connell, Warren Siegel, George Sterman, and Ilmo Sung for useful comments. YH would like to thank Mark Wise for the invitation as visiting scholar at Caltech.
\appendix

\section{ Review of BLG and ABJM\label{APPA}}

The BLG theory is a three-dimensional superconformal Chern-Simons theory with classical $OSp(8|4)$ symmetry. The matter content consists of eight scalars and 8 Majorana fermions. It can be written as an $SO(4)$ gauge theory with matter in the fundamental representation of $SO(4)$:

\[
\mathcal{L}= -\frac{1}{2}D_{\mu}X^{Ia}D^{\mu}X^{Ia}+\frac{i}{2}\bar{\psi}^{a}\gamma^{\mu}D_{\mu}\psi^{a}\]
\[+\frac{1}{8}\epsilon^{\mu\nu\lambda}\epsilon^{abcd}\left(  A_{\mu ab}\partial_{\nu}A_{\lambda cd}+\frac{2}{3}g A_{\mu ab}A_{\nu ce}A_{\lambda ed} \right )\]
\[-\frac{i}{4}g^{2}\epsilon^{abcd}\bar{\psi}_{a}\Gamma^{IJ}\psi_{b}X_{c}^{I}X_{d}^{J}-\frac{1}{12}g^{4}\epsilon^{abcd}\epsilon^{aefh}X^{Ib}X^{Jc}X^{Kd}X^{Ie}X^{Jf}X^{Kh}\]
where $I=1,...,8$, $a=1,...,4$ are $SO(4)$ indices, $D_{\mu}X_{a}^{I}=\partial_{\mu}X_{a}^{I}+gA_{\mu ab}X_{b}^{I}$, $g=\sqrt{\pi/k}$, and $k$ is an integer called the level. Note that $\Gamma^{IJ}=\frac{1}{2}\left(\Gamma^{I}\left(\Gamma^{J}\right)^{T}-\Gamma^{J}\left(\Gamma^{I}\right)^{T}\right)$ (we use the conventions of \cite{Bandres:2008vf}). Furthermore, $\gamma^\mu$ are the 3d Dirac matrices and $\Gamma^I$ are the Clebsch-Gordan coefficients relating the three different 8d representations of the $Spin(8)$. In particular, the scalars transform in the $8_v$ representation and the fermions transform in the $8_s$ representation (although their $Spin(8)$ indices are not written explicitly). Note that the theory is weakly coupled when $k \gg 1$ and that the gauge field transforms in the adjoint representation of $SO(4)$, i.e. $A_{\mu ab}=-A_{\mu ba}$. Also note the appearance of the four-index invariant tensor $\epsilon^{abcd}$. This tensor can be interpreted as the structure constant $f^{abcd}$ for a so-called 3-algebra which obeys a generalization of the Jacobi identity:
\begin{equation}
f_{\,\,\,\,\,\,\,\,\,\,\, e}^{[abc}f_{\,\,\,\,\,\,\,\,\,\,\, g}^{d]ef}=0
\label{fund}
\end{equation}
where indices are raised and lowered using the three-algebra metric $h_{ab}=\delta_{ab}$. This formula is referred to as the fundamental identity. It has been proven that if $h_{ab}$ is positive-definite, then the only totally antisymmetric structure constant which satisfies the fundamental identity is $\epsilon^{abcd}$ \cite{Gauntlett:2008uf}. In this sense, the BLG theory is unique.

Three-algebras are a generic feature of superconformal Chern-Simons theories. In particular, the fundamental identity follows from the closure of supersymmetry. Although there is only one three-algebra with positive-definite metric that gives $\mathcal{N}=8$ supersymmetry, if one considers three-dimensional theories with lower amounts supersymmetry, one finds a lot more possibilities. For example, there is an infinite family of three algebras corresponding to the ABJM theory, which has $\mathcal{N}=6$ supersymmetry and gauge group $U(N) \times U(N)$. In this case, the four-index structure constant is complex, not totally antisymmetric, and satisfies a slightly more general version of the fundamental identity \cite{Bagger:2008se}.

Since $SO(4) \sim SU(2) \times SU(2)$, the BLG theory can also be expressed as an $SU(2) \times SU(2)$ gauge theory, where the matter transforms in the bi-fundamental representation of the gauge group \cite{raamsdonk}:
\[
\mathcal{L}={\normalcolor Tr}[-D_{\mu}X^{I\dagger}D^{\mu}X^{I}+i\bar{\psi}\gamma^{\mu}D_{\mu}\psi\]
\[+\frac{1}{4}\epsilon^{\mu\nu\lambda}\left(A_{\mu}\partial_{\nu}A_{\lambda} +\frac{2i}{3}g A_{\mu}A_{\nu}A_{\lambda}-\hat{A}_{\mu}\partial_{\nu}\hat{A}_{\lambda} - \frac{2i}{3}g \hat{A}_{\mu}\hat{A}_{\nu}\hat{A}_{\lambda}\right)\]
\[-\frac{2i}{3}g^{2}\bar{\psi}\Gamma^{IJ}\left(X^{I}X^{J\dagger}\psi+X^{J}\psi^{\dagger}X^{I}+\psi X^{I\dagger}X^{J}\right)-\frac{8}{3}g^{4}X^{[I}X^{J\dagger}X^{K]}X^{K\dagger}X^{J}X^{I\dagger}]\]
where $D_{\mu}X^{I}=\partial_{\mu}X^{I}+ig\left(A_{\mu}X^{I}-X^{I}\hat{A}_{\mu}\right)$. Note that the gauge fields $A_{\mu}$ and $\hat{A}_{\mu}$ are associated with each $SU(2)$ and appear with opposite signs in the action. For this reason, the BLG model is referred as a twisted Chern-Simons theory. The fields can be expanded in terms of generators as follows:\[
X^{I}=X^{Ia}\tilde{T}^{a},\,\,\, A_{\mu}=A_{\mu}^{j}T^{j},\,\,\,\hat{A}_{\mu}=\hat{A}_{\mu}^{j}T^{j}\]
where $a=1,2,3,4$, and $j=1,2,3$. Note that there's a similar expression for the spinors. Explicit formulas for the generators are
\[
T^{1}=\left(\begin{array}{cc}
0 & 1\\
1 & 0\end{array}\right),\,\,\, T^{2}=\left(\begin{array}{cc}
0 & -i\\
i & 0\end{array}\right),\,\,\, T^{3}=\left(\begin{array}{cc}
1 & 0\\
0 & -1\end{array}\right)\]
 \[
\tilde{T}^{j}=\frac{i}{2}T^{j},\,\,\,\tilde{T}^{4}=\frac{1}{2}\left(\begin{array}{cc}
1 & 0\\
0 & 1\end{array}\right).\]
Using this formulation of the BLG theory, one can define color-ordering in scattering amplitudes, as demonstrated in the next Appendix.

The ABJM theory can also be written as a three-dimensional twisted Chern-Simons theory with bi-fundamental matter. Since it shares several features in common with the BLG theory, we will only describe the action schematically. For more details see \cite{Bandres:2008ry, Benna:2008zy}. The field content consists of four complex scalars $Z^{A}$ and four Dirac fermions $\psi_A$ transforming in the bi-fundamental representation of $U(N)\times U(N)$ (with $A$ running from 1 to 4), as well as two $U(N)$ gauge fields $A_{\mu}$ and $\hat{A}_{\mu}$. The matter fields transform in the fundamental representation of the R-symmetry group $SU(4)$ and their adjoints transform in the anti-fundamental representation of $SU(4)$. For $k^5 \ll N$, the theory is dual to M-theory on $AdS_4 \times S^7/ \mathbb{Z}_k $, and for $k\ll N\ll k^{5}$, it is dual to type IIA string theory on $AdS_{4}\times CP^3$. For $k=1,2$, the theory is conjectured to have $\mathcal{N}=8$ supersymmetry, unlike the BLG theory which has maximal supersymmetry for all values of the level. It is interesting to note however, that one can gauge-fix the ABJM theory to have $SU(N) \times SU(N) \times \mathbb{Z}_k $ gauge symmetry, where $Z^{I}\rightarrow e^{2\pi i/k}Z^{I}$ under the discreet gauge group. Furthermore, if one neglects the discreet gauge group and sets $N=2$, it can be shown that one retrieves the BLG theory.
\section{ BLG 4-pt Using Bi-fundamental Notation\label{APPB}}

The interaction terms needed to compute the tree-level 4-pt scalar
amplitude are
\begin{equation}
\mathcal{L}_{int}=ig{\normalcolor Tr}\left(X^{I\dagger}A_{\mu}\partial_{\mu}X^{I}-\partial_{\mu}X^{I\dagger}A_{\mu}X^{I}+\partial_{\mu}X^{I\dagger}X^{I}\hat{A}_{\mu}-X^{I\dagger}\partial_{\mu}X^{I}\hat{A}_{\mu} \right)
\label{eq}
\end{equation}
where $g=\sqrt{\pi/k}$ and $k$ is the level.

In field theories where the fields are matrices, it is convenient
to draw Feynman diagrams using double-line notation which indicates
how the matrix indices are contracted. At the same time, these diagrams can be factorized
into a color factor and a color-ordered Feynman diagram which can
be represented using single-line notation. The color-ordered propagators
for the gauge fields are depicted in Figure \ref{F1} and are given by \[
\pm\left(\epsilon_{\mu\nu\lambda}p^{\lambda}-i\xi\frac{p_{\mu}p_{\nu}}{4p^{2}}\right)\frac{2}{p^{2}}\]
where $+/-$ corresponds to $A_{\mu}/\hat{A}_{\mu}$ and $\xi$ is
a gauge-fixing parameter. The color-ordered Feynman diagrams associated
with the 3-point interactions in equation Eq \ref{eq} are depicted
in Figure \ref{F2}. They are given by \[
\pm ig\left(p_{a}-p_{c}\right)_{\mu}\delta^{IJ}{\normalcolor }\]
where the $+$ corresponds to the $A_{\mu}$ vertex and $-$  corresponds to the $\hat{A}_{\mu}$ vertex.
Note that all momenta are taken to be outgoing. The double-line version
of the 3-point vertices is illustrated in Figure \ref{F3}.
\begin{figure}
\center
\includegraphics[height=14mm]{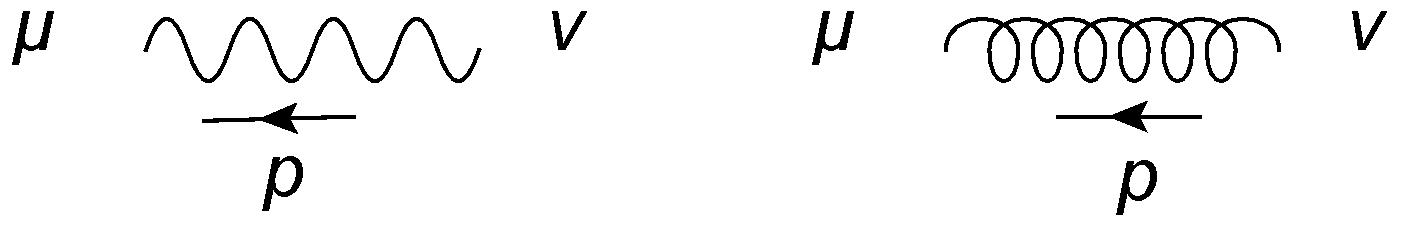}
\caption{Color-ordered propagators. The gauge field $A_{\mu}$ is represented by a wavy line and the gauge field $\hat{A}_{\mu}$ is represented by a curly line.}
\label{F1}
\end{figure}

\begin{figure}
\center
\includegraphics[height=30mm]{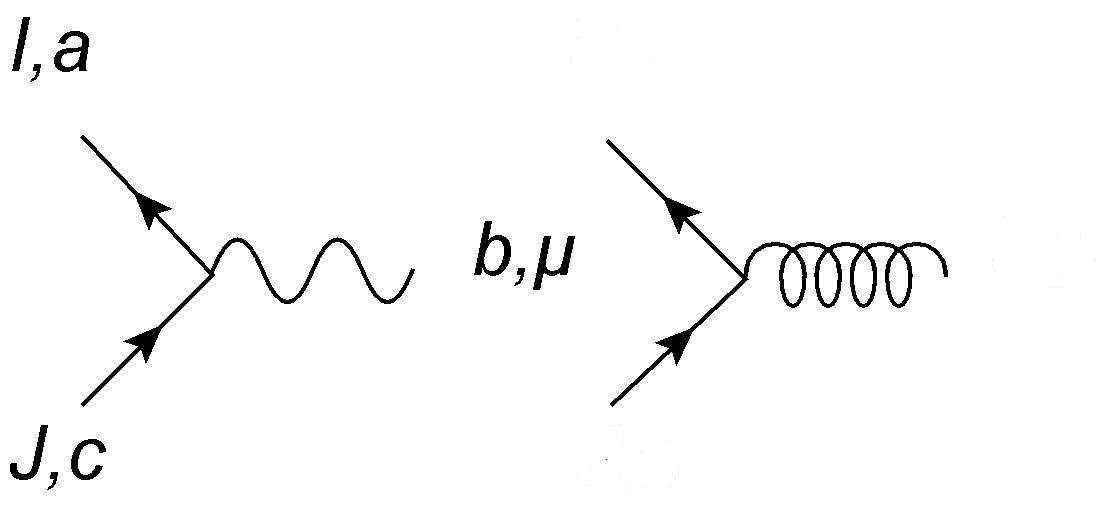}
\caption{Color-ordered 3-point vertices. The first diagram corresponds to the first two terms in Eq (B.1) and the second diagram corresponds to the third and fourth terms in Eq (B.1).}
\label{F2}
\end{figure}

\begin{figure}[tb]
\center
\includegraphics[height=30mm]{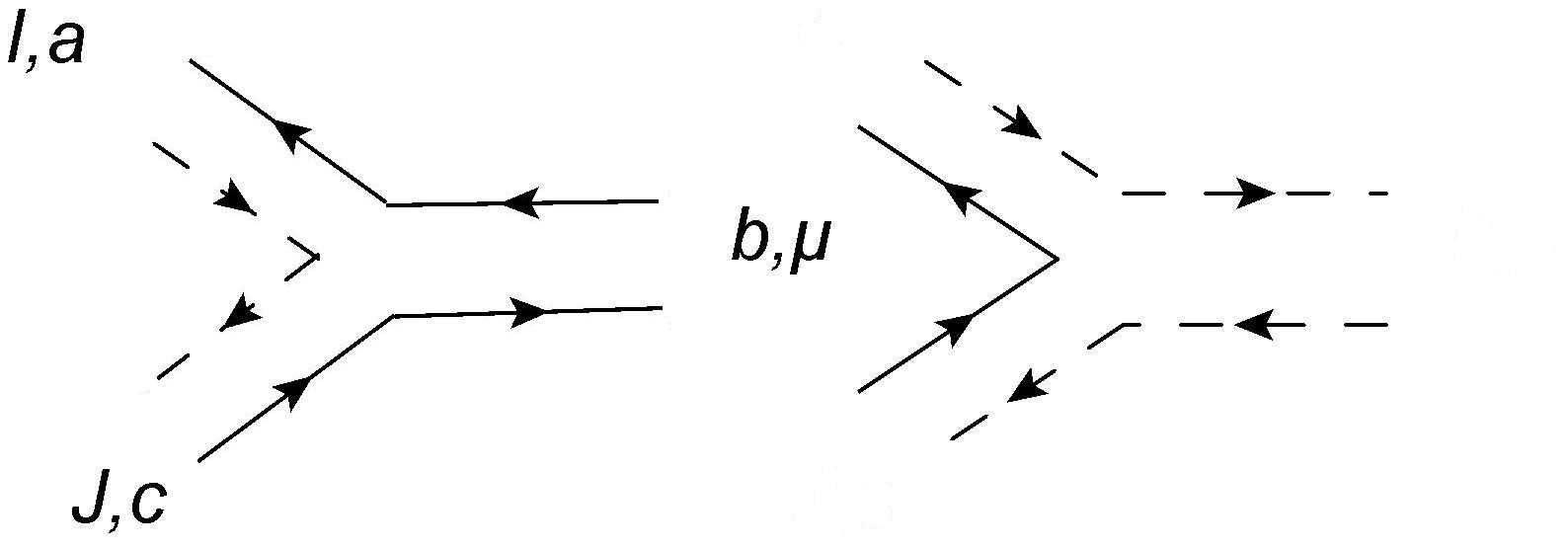}
\caption{3-point vertices in double line notation. Note that there are two types of lines since the gauge group is $SU(2)\times SU(2)$. The color factor for the first diagram is $\tilde{T}^{c}\tilde{T}^{a\dagger}$ and the color factor for the second diagram is $\tilde{T}^{a\dagger}\tilde{T}^{c}$.}
\label{F3}
\end{figure}

\begin{figure}[tb]
\center
\includegraphics [height=60mm]{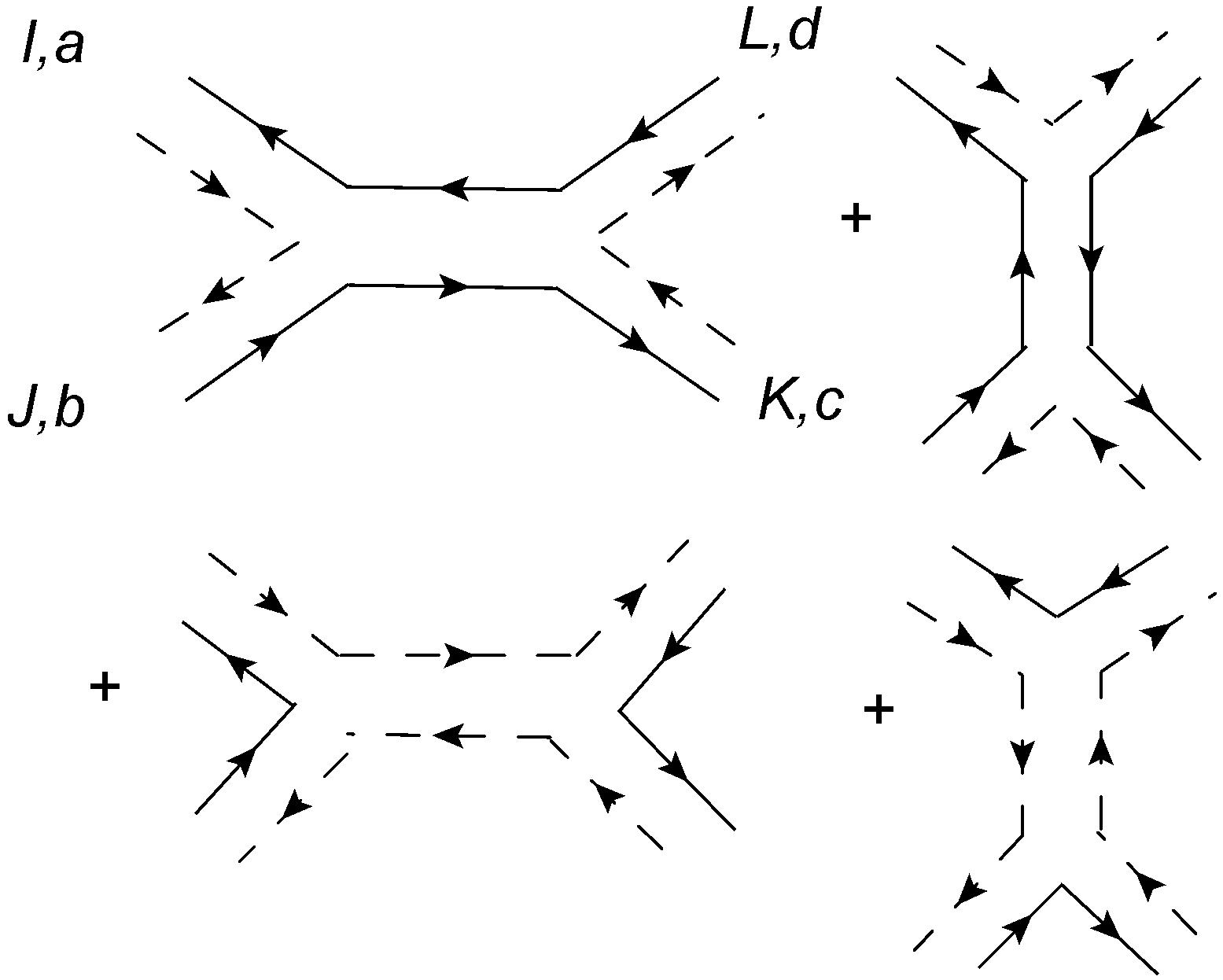}
\caption{Planar diagrams contributing to tree-level 4-pt scalar amplitude. Non-cyclic permutations of the external legs correspond to $\left(Jb,Kc,Ld\right)\rightarrow\left(Kc,Ld,Jb,\right),\left(Ld,Jb,Kc,\right),\left(Kc,Jb,Ld\right),\left(Jb,Ld,Kc\right),\left(Ld,Kc,Jb\right)$.}
\label{F4}
\end{figure}

To compute the tree-level scattering amplitude, we only need to consider
the planar diagrams depicted in Figure \ref{F4}. The full amplitude is then
given performing non-cyclic permutations of the external legs. In total, there are 24 diagrams. Furthermore,
each diagram factorizes into a trace and a color-ordered
Feynman diagram which can be evaluated using the color-ordered Feynman
rules described above. For example, the first diagram
in Figure \ref{F4} is given by

\[
4g^{2}{\normalcolor Tr}\left(\tilde{T}^{b}\tilde{T}^{a\dagger}\tilde{T}^{d}\tilde{T}^{c\dagger}\right)\delta^{IJ}\delta^{KL}\epsilon_{\mu\nu\lambda}\frac{p_{a}^{\mu}p_{b}^{\nu}\left(p_{c}-p_{d}\right)^{\lambda}}{\left(p_{a}+p_{b}\right)^{2}}\]
and the third diagram in Figure 4 is given by

\[
-4g^{2}{\normalcolor Tr}\left(\tilde{T}^{a\dagger}\tilde{T}^{b}\tilde{T}^{c\dagger}\tilde{T}^{d}\right)\delta^{IJ}\delta^{KL}\epsilon_{\mu\nu\lambda}\frac{p_{a}^{\mu}p_{b}^{\nu}\left(p_{c}-p_{d}\right)^{\lambda}}{\left(p_{a}+p_{b}\right)^{2}}.\]
Performing the non-cyclic permutations and noting that\[
{\normalcolor Tr}\left(\tilde{T}^{a}\tilde{T}^{b\dagger}\tilde{T}^{c}\tilde{T}^{d\dagger}\right)=-{\normalcolor Tr}\left(\tilde{T}^{a\dagger}\tilde{T}^{b}\tilde{T}^{c\dagger}\tilde{T}^{d}\right)=\frac{1}{8}\epsilon^{abcd}{\normalcolor ,}\]
one finds that the 4-pt amplitude is given by
\[\mathcal{A}_{4}=4g^{2}\delta^{3}(P)\epsilon_{abcd}\epsilon^{\mu\nu\lambda}p_{a\mu}\left[\delta^{IJ}\delta^{KL}\frac{p_{b\nu}\left(p_{c}-p_{d}\right)_{\lambda}}{\left(p_{a}+p_{b}\right)^{2}}+\delta^{IL}\delta^{KJ}\frac{p_{d\nu}\left(p_{b}-p_{c}\right)_{\lambda}}{\left(p_{a}+p_{d}\right)^{2}}+\delta^{IK}\delta^{JL}\frac{p_{c\nu}\left(p_{d}-p_{b}\right)_{\lambda}}{\left(p_{a}+p_{c}\right)^{2}}\right]\]
which matches the result we obtained using $SO(4)$ notation if one notes that
$\epsilon^{\mu\nu\lambda}p_{a\mu}p_{b\nu}\left(p_{c}-p_{d}\right)_{\lambda}=\epsilon^{\mu\nu\lambda}p_{a\mu}p_{d\nu}\left(p_{b}-p_{c}\right)_{\lambda}=\epsilon^{\mu\nu\lambda}p_{a\mu}p_{c\nu}\left(p_{d}-p_{b}\right)_{\lambda}$.

\section{SU(2) Spinors for 6D 3-pt Amplitudes\label{APPC}}
It is well known that for the 3-pt amplitude, momentum conservation and on-shell conditions lead to vanishing Lorentz invariants:
$$p_1+p_2+p_3=0\rightarrow p_1\cdot p_2=p_1\cdot p_3=p_2\cdot p_3=0.$$
As discussed in \cite{Cheung:2009dc}, this leads to a vanishing determinant of the inner product between a chiral and an anti-chiral
spinor in six dimensions
$$(p_i)^{AB}(p_j)_{AB}=0\rightarrow \det\langle i_a| j_{\dot{a}}]=0.$$
Since the 2$\times$2 matrix $\langle i_a|j_{\dot{a}}]$ has rank 1, it can be written in terms of $SU(2)$ spinors,
i.e. $\langle i|j]_{a{\dot{a}}}=u_{ia}\tilde{u}_{j\dot{a}}$. Consistently defining the SU(2) spinors for all Lorentz invariants gives
\eqa
\nonumber\langle1_a|2_{\dot{b}}]=u_{1a}\tilde{u}_{2\dot{b}},\;\langle2_a|1_{\dot{b}}]=-u_{2a}\tilde{u}_{1\dot{b}}\\
\nonumber\langle2_a|3_{\dot{b}}]=u_{2a}\tilde{u}_{3\dot{b}},\;\langle1_a|3_{\dot{b}}]=-u_{1a}\tilde{u}_{3\dot{b}}\\
\nonumber\langle3_a|1_{\dot{b}}]=u_{3a}\tilde{u}_{1\dot{b}},\;\langle3_a|2_{\dot{b}}]=-u_{3a}\tilde{u}_{2\dot{b}}.\\
\label{not}
\eqae
One important property of these SU(2) spinors can be derived from momentum conservation,
\eqa
\nonumber\left[\lambda_1\cdot(p_1+p_2+p_3)\right]_A=0&\rightarrow& \langle 1_a|2^{\dot{b}}][2_{\dot{b}}|_A+\langle 1_a|3^{\dot{c}}][3_{\dot{c}}|_A=0\\
&\rightarrow& \tilde{u}_2^{\dot{c}}[2_{\dot{c}}|_A=\tilde{u}_3^{\dot{c}}[3_{\dot{c}}|_A=\tilde{u}_1^{\dot{c}}[1_{\dot{c}}|_A.
\label{law1}
\eqae
Another useful identity is $|\tilde{u}_K\cdot \tilde{u}_P|=\sqrt{-s_{14}}$, which holds in the collinear limit of the four point kinematics depicted in fig.(\ref{schannel}). This can be proven as follows:
\footnote{We thank Donal O'Connell for discussion of the proof of this identity.}
Consider the following object:
\eqa
\nonumber \langle 4_{a}|p_3p_{1} |4_{\dot{a}}]&=& u_{4a}\tilde{u}_3^{\dot{d}}[ 3_{\dot{d}}|p_{1}|4_{\dot{a}}]\\
&=& u_{4a}\tilde{u}^{\dot{d}}_{4}[ 4_{\dot{d}}|p_{1}|4_{\dot{a}}]=u_{4a}\tilde{u}_{4\dot{a}}s_{14}.
\label{claytondown}
\eqae
On the other hand, one can also deduce
\eqa
\nonumber \langle 4_{a}|p_3p_{1} |4_{\dot{a}}]&=& u_{4a}\tilde{u}^{\dot{d}}_{K}[ K_{\dot{d}}|p_{1}|4_{\dot{a}}]= iu_{4a}\tilde{u}^{\dot{d}}_{K}[ P_{\dot{d}}|p_{1}|4_{\dot{a}}]\\
\nonumber &=& iu_{4a}(\tilde{u}_{K}\cdot\tilde{u}_{P})u^b_{1}\langle 1_{b}|4_{\dot{a}}]=iu_{4a}(\tilde{u}_{K}\cdot\tilde{u}_{P})u^b_{P}\langle P_{b}|4_{\dot{a}}]\\
&=&  -u_{4a}\tilde{u}_{4\dot{a}}(\tilde{u}_{K}\cdot\tilde{u}_{P})^2.
\label{claytonup}
\eqae
Comparing eqs.(\ref{claytondown}) and (\ref{claytonup}) gives
\eq
|\tilde{u}_{K}\cdot\tilde{u}_{P}|=\sqrt{-s_{14}}.
\label{wordyo}
\eqe


\end{document}